\documentclass[prd,preprint,showpacs,preprintnumbers]{revtex4}
\usepackage{graphicx}
\def \beq{\begin{equation}}
\def \eeq{\end{equation}}
\def \beqa{\begin{eqnarray}}
\def \eeqa{\end{eqnarray}}
\def\E#1{\times10^{-#1}}
\def\e#1{10^{-#1}}
\def \ie{{\sl i.e.\/}}
\def \sn{{\rm sn}}
\def \cn{{\rm cn}}

\def \machine{Fujitsu VPP5000}
\def \ncg{{N_{CG}}}
\def \nor{{N_O}}
\def \ndf{{N_D}}
\def \tr{{\rm Tr}\,}
\def \ehm{\epsilon_{\scriptscriptstyle H}}
\def \enm{\epsilon_{\scriptscriptstyle N}}
\def \ecg{\epsilon_{\scriptscriptstyle CG}}
\def \ecc{\epsilon_{\scriptscriptstyle CC}}
\def \ehf{\epsilon_{\scriptscriptstyle 1/2}}
\def \dgw{\epsilon'_{\scriptscriptstyle GW}}
\def \dhm{\epsilon'_{\scriptscriptstyle H}}

\def \dcc{\epsilon'_{\scriptscriptstyle CC}}
\def \dhf{\epsilon'_{\scriptscriptstyle 1/2}}
\def \egw{\epsilon_{\scriptscriptstyle GW}}

\def \zhm{Z_{H}}
\def \znm{Z_{N}}
\def \zcc{Z_{CC}}
\def \zhf{Z_{1/2}}
\def \zgw{Z_{GW}}
\begin{document}

\title{Speed and Adaptability of Overlap Fermion Algorithms}
\author{Rajiv V.\ \surname{Gavai}}
\email{gavai@tifr.res.in}
\author{Sourendu \surname{Gupta}}
\email{sgupta@tifr.res.in}
\affiliation{Department of Theoretical Physics, Tata Institute of Fundamental
         Research,\\ Homi Bhabha Road, Mumbai 400005, India.}
\author{Robert \surname{Lacaze}}
\email{lacaze@spht.saclay.cea.fr}
\affiliation{Service de Physique Theorique, CEA Saclay,\\
         F-91191 Gif-sur-Yvette Cedex, France.}
\begin{abstract}
We compare the efficiency of four different algorithms to compute the overlap
Dirac operator, both for the speed, {\sl i.e.\/}, time required to reach a
desired numerical accuracy, and for the adaptability, {\sl i.e.\/}, the scaling
of speed with the condition number of the (square of the) Wilson Dirac
operator.  Although orthogonal polynomial expansions give good speeds at
moderate condition number, they are highly non-adaptable. One of the rational
function expansions, the Zolotarev approximation, is the fastest and is
adaptable.  The conjugate gradient approximation is adaptable, self-tuning, and
nearly as fast as the ZA.
\end{abstract}
\pacs{11.15.Ha, 12.38.Mh\hfill TIFR/TH/02-23, t02/091, hep-lat/0207005}
\maketitle

\section{Introduction}

A lack of consistent definition of chiral fermions on the lattice has hampered
definitive and convincing investigations of chiral aspects of quantum
chromodynamics (QCD) until now. Thus important physics issues, such as the
spontaneous breaking of the chiral symmetry at low temperatures and its
restoration at finite temperature, have remained hostages to technical questions
such as the fine-tuning of the bare quark mass (Wilson fermions) or the precise
number of massless flavours (staggered fermions).  Recent developments in
defining exact chiral symmetry on the lattice have therefore created exciting
prospects of studying an enormous amount of physics in a cleaner manner from
first principles.  However, the corresponding Dirac operators are much more
complicated.  Without good control of the algorithms needed to deal with
them, one is unlikely to derive the full benefit of their better chiral
properties.  Our goal in this paper is to evaluate the efficiency of the most
widely used, or most promising, algorithms. By efficiency we mean both the speed
of the algorithm, which is measured by the computer time required to achieve a
certain accuracy in the solution, and the adaptability, which is measured by
how the speed scales as the problem becomes harder.
This study is made for various values of the required accuracy along with the
corresponding analysis on the accuracy obtained for the expected properties
of the resulting Dirac operator such as the Ginsparg-Wilson relation, the 
central circle relation, $\gamma_5$ hermiticity or normality.
In particular, we have observed that these properties can be satisfied
accurately only if the sign computation of the Wilson Dirac operator
has high enough precision.

\subsection{The overlap Dirac operator}

One version of chiral fermions on the lattice is the overlap formalism.
The overlap Dirac operator ($D$) is defined \cite{neu2} in terms of the
Wilson-Dirac operator ($D_w$) by the relation
\beq
   D = 1 + D_w (D_w^\dag D_w)^{-1/2}.
\label{overlap}\eeq
In this paper we shall use the shorthand notation
\beq
 M=D_w^\dag D_w~.
\label{defM}\eeq
The Wilson-Dirac operator $D_w$ (for lattice spacing $a =1$) is given by
\beq
   D_w = \frac{1}{2} \sum_\mu \big[ \gamma_\mu \big( \partial_\mu
+ \partial_\mu^* \big) - \partial_\mu \partial_\mu^* \big] + m ~,~
\label{wilsond}\eeq
where $\partial_\mu$ and $\partial_\mu^*$ are (gauge covariant) forward and backward
difference operators respectively.  It has been shown that as long as the mass $m$ is
in the range $-2<m<0$, the above overlap Dirac operator is well-defined, and corresponds
to a single massless fermion.  Furthermore, it satisfies the Ginsparg-Wilson
relation \cite{gw}
\beq
   \gamma_5 D + D\gamma_5 = D \gamma_5 D,
\label{gwrel}\eeq
which leads to a good definition of chirality on the lattice and has been
shown to correspond to an exact chiral symmetry on the lattice.

The overlap Dirac operator, $D$, enjoys many nice properties in addition
to the Ginsparg-Wilson relation in Eq.\ (\ref{gwrel}). In particular, it
satisfies $\gamma_5$-Hermiticity---
\beq
   D^\dag = \gamma_5D\gamma_5.
\label{g5her}\eeq
Together with the Ginsparg-Wilson relation, this implies that $D$ is
normal, {\sl i.e.\/}, 
\beq
   [D,D^\dag]=0.
\label{normal}\eeq
Normality clearly means that $D$ and $D^\dag$ have the same eigenvectors.
Eqs.\ (\ref{gwrel},\ref{g5her}) also imply
\beq
   D+D^\dag = D^\dag D,
\label{circle}\eeq
and hence the eigenvalues of $D$ and $D^\dag$ lie on the unit circle
centered at unity on the real line, implying that $ D -1$ is unitary. 
We define measures of numerical
errors on each of these quantities, and relations between them in Section
\ref{sc.errors}.

\subsection{Numerical algorithms}

All computations of hadronic correlators involve the determination
of the fermion propagator $D^{-1}$, and need a nested series of two
matrix iterations for their evaluation, since each step in the numerical
inversion of $D$ involves the evaluation of $M^{-1/2}$.  This squaring
of effort makes a study of QCD with overlap quarks very expensive.

This problem defines for us the properties that an efficient algorithm to
deal with $M^{-1/2}$ must have. First, it should achieve a given desired
accuracy as quickly as possible. The need for accuracy is clear:  the
accuracy to which the Ginsparg-Wilson relation in Eq.\ (\ref{gwrel})
is satisfied depends on the accuracy achieved in the computation of
$M^{-1/2}$.  The second, and equally important, requirement is that
the method should adapt itself easily to matrices with widely different
condition numbers---
\beq
   \kappa = \frac{\lambda_{\max}}{\lambda_{\min}},
\label{cond}\eeq
where $\lambda_{\min}$ and $\lambda_{\max}$ are, respectively, the
minimum and the maximum eigenvalues of $M$. Adaptability is needed
because in QCD applications the eigenvalue spectrum of $M$ can fluctuate
over many orders of magnitude from one configuration to the next. Since
the condition number on a configuration is {\sl a priori\/} unknown,
a method with low adaptability will end up either being inefficient or
inaccurate or even both.   Algorithms, which are adaptable in principle, 
may require tuning of parameters by hand, or they may incorporate a procedure 
for self-tuning.  Clearly, self-tuning algorithms are the ones that can best
deal with fluctuating condition numbers in any real situation.

In this paper we examine the speed and adaptability of several
different algorithms to compute the inverse square root, $M^{-1/2}$,
of Hermitean matrices (in our applications the eigenvalues of $M$
are non-negative ) acting on a vector.
Several algorithms for this have been proposed in the literature.

We do not consider the first algorithm to be proposed, since
this requires a matrix inversion to be performed at each step of
an iteration to determine $M^{-1/2}$ \cite{chiu}. Later algorithms
are more efficient. These fall into two classes--- expansions of
$1/\sqrt z$ in appropriate classes of functions (rational functions
\cite{neu} or orthogonal polynomials \cite{hepilu}), and iterative
methods \cite{borici}.  We have analyzed four derived algorithms, namely
the Optimized Rational Approximation (ORA) \cite{nara}, the Zolotarev
Approximation (ZA, which is also a rational expansion) \cite{wupp,ch},
the Chebychev Approximation (CA, a polynomial expansion) \cite{hepile}
and the Conjugate Gradient Approximation (CGA, an iterative method)
\cite{ragula}.  We find that an expansion in Chebychev polynomials is the
fastest when $\kappa$ is moderately large, but it suffers from various
instabilities including a lack of adaptability. Rational expansions cure
many of the instabilities of polynomial expansions; indeed the ZA is the
fastest and is adaptable but not self-tuning.
An iterative method is the only fully self-tuned algorithm,
and it turns out to be reasonable also from the point of view of speed.

\subsection{Algorithmic costs and adaptability}

We make two different estimates of the cost of each algorithm. The
complexity, $\cal C$, counts the number of arithmetic operations
required to achieve the solution to the problem and is a measure of
speed independent of the specific machine on which the algorithm is
implemented. The spatial complexity, $\cal S$, is the memory requirement
for the problem. While timing runs on particular machines on chosen test
configurations are instructive, the scaling of speed for each algorithm
with physical and algorithmic parameters is provided by our estimates of
$\cal C$.

Our estimate of the adaptability, $\cal A$, of each algorithm is the
following. If the scalar algorithm for $1/\sqrt z$ is tuned to have
maximum relative error $\varepsilon$ in the range $[z_{\min},z_{\max}]$,
then we find the smallest range $[z'_{\min},z'_{\max}]$ where the
error is at most $10\varepsilon$. Note that the second interval cannot be
smaller than the first. In terms of these quantities we define
\beq
   {\cal A} = \frac{\log(z'_{\max}/z'_{\min})}{\log(z_{\max}/z_{\min})}
      -1.
\label{adapt}\eeq
The least adaptable algorithms have small values of $\cal A$
( ${\cal A}>0$ ). $\cal A$ is a
measure of the relative accuracy achieved in a fixed CPU time for the
same algorithm running on two different configurations with condition
numbers $\kappa=z_{\max}/z_{\min}$ and $\kappa'=z'_{\max}/z'_{\min}$. In
conjunction with estimates of $\cal C$, it also contains information
about the scaling of CPU time required to achieve the same accuracy on
the two configurations.

\subsection{Numerical tests}

Our numerical tests were performed with three typical $SU(3)$ gauge
configurations on a $4\times12^3$ lattice at $\beta=5.80$
({\sl i.e.\/}, $T=1.25T_c$).  The configuration A had eigenvalues of $M$
in the range $[0.032,32]$ so that $\kappa=10^3$. The configuration
B had eigenvalues in the range $[7.2\times10^{-5},32]$, giving
$\kappa=4.4\times10^5$.  Finally, configuration C had eigenvalues in
the range $[8.9\times10^{-9},32]$ and hence $\kappa=3.6\times10^9$.
Configuration A is one of the easiest configuration we found in our
simulations, and there were several configurations with $\kappa$ of
order $10^5$--$10^9$ in the data set we worked with in \cite{ragula}.
If there is a single scale in the level spacing of the eigenvalues of
$M$, then we expect $\kappa={\cal O}(V)\approx7\times10^3$ on our test
configurations. We conclude that A is indeed a little easier than the
generic configuration and B and C are successively harder.  The CPU
times we quote in our tables are obtained on a \machine, which is a
vector computer. Our computations ran on this machine at a speed of
around 4.1 Gigaflops.

In sections \ref{sc.ca}--\ref{sc.cga} we describe and analyze the four
algorithms and also present data on precision and time measurements on
these three test configurations. In this work we have not investigated
the performance of the algorithms with deflation (explicit subtraction)
of some eigenvectors. However, we do remark on the precision required
for deflation and the propagation of errors due to such a subtraction.
Section \ref{sc.comp} contains a comparison of the numerical results
and our conclusions.

\section{Errors}\label{sc.errors}

In general, every numerical method to compute $M^{-1/2}$ constructs an 
operator $L$ which applied to a vector $\Phi$ gives the vector

\beq
   X=L[\Phi] \qquad{\rm with}\qquad L[\Phi] = M^{-1/2}\Phi+E[\Phi],
\label{approx}\eeq
where $E$ is the error in the
approximation, $L$, to the matrix $M^{-1/2}$.
Typically, these operators $L$ and $E$ are not matrices because the algorithms
can introduce a dependence on $\Phi$ which is not linear.
The error $E[\Phi]$ on the computation of $M^{-1/2}$  leads to the violation
of the properties in Eqs. (\ref{gwrel}-\ref{circle}).  In our numerical 
tests we have investigated five measures of the accuracy of the algorithms 
through norms of the following operators---

\beqa
\nonumber
   \zhf = ML^2 - 1, \quad&
   \zgw = D\gamma_5+\gamma_5D-D\gamma_5D, \quad&
   \znm = DD^\dag-D^\dag D,\\
   \zhm = D^\dag-\gamma_5 D\gamma_5, \quad&
   \zcc = D+D^\dag-D^\dag D,\quad&
\label{analyse}\eeqa
where $D=1+D_wL$ and $D^\dag=1+LD_w^\dag$.  Each of these operators is zero
when $E[\Phi]$ = 0. With Gaussian random vectors $\Phi$, we have
measured the deviations from this exact value through
\beq
   \epsilon_i = |Z_i\Phi|/|\Phi| \qquad{\rm and}\qquad
   \epsilon'_i = \Phi^\dag Z_i\Phi/|\Phi|^2.
\label{errors}\eeq
Here, and later, we use the notation $|v|=\sqrt{v^\dag v}$ for any complex
vector $v$. Note that $\epsilon_i$ are real and non-negative whereas
$\epsilon'_i$ are complex in general.

\subsection{Polynomial approximation}

The polynomial approximation consists of writing
\beq
   L = \sum_{i=1}^{\nor}b_i M^i ,
\label{iter}\eeq
where $b_i$ are constants.  It is clear that
both $L$ and $E$ are matrices in this case and 
\beq
[L,M] = 0 .
\label{lmcom}\eeq
Since in numerical implementations of $D_w$, $\gamma_5$-Hermiticity is
accurate to machine precision, {\sl i.e.\/}, $|(D_w^\dag -
\gamma_5D_w\gamma_5)\Phi| = 0$, one has the following relation:
\beq
\gamma_5D_wM^n\gamma_5=M^nD_w^\dag . 
\label{polher}\eeq
Its direct consequence is 
\beq
  \zhm = LD_w^\dag - \gamma_5D_wL\gamma_5 = 0 .
\label{gherm}\eeq

Using Eqs.(\ref{lmcom}-\ref{polher}), one can easily obtain
the following relations between the various $Z$'s :
\beqa
\nonumber
   \zcc &=& - \zhf,\\
\nonumber
   \zgw &=& \gamma_5 \zcc,\\
   \znm &=& \zcc -\gamma_5 \zcc \gamma_5. 
\label{zrelations}\eeqa
As a consequence, 
\beqa
\nonumber
   |\dhf| &=&|\dcc|,\\
\nonumber
   \ehf &=& \ecc=\egw, \\
   \dhm &=& \ehm=0.
\label{rel1}\eeqa

Expanding $\zhf=ML^2-1=\zhf^\dag$ in powers of $E$ and retaining only
the leading order terms, we find that $\zhf^\dag\zhf=4E^2M$ and
$\zhf=2EM^{1/2}$.  Defining the averages of $\epsilon_i$ and
$\epsilon'_i$ over an ensemble of $\Phi$ as 
$\overline{ \epsilon_i^2} = \tr Z_i^\dag Z_i$ and 
$\overline{ \epsilon'_i} = \tr Z_i$, one  obtains,
\beqa
\nonumber
   \overline{\ehf^2}=\overline{\ecc^2}=\overline{\egw^2}=4\int d\lambda\rho(\lambda)\lambda \epsilon^2(\lambda),\\
\nonumber
   \overline{\dhf}=-\overline{\dcc}=2\int d\lambda\rho(\lambda)\sqrt\lambda \epsilon(\lambda),\\
\overline{\dgw}=2\int d\lambda\Delta\rho(\lambda)\sqrt\lambda\epsilon(\lambda),
\label{ehfsq3}\eeqa
where $\rho(\lambda)$ is the density of eigenvalues of $M$, 
$\epsilon(\lambda)$ is the error in the approximation and $\Delta\rho(\lambda)=\rho_+(\lambda)-\rho_-(\lambda)$, the
difference between the spectral densities (of $M$) in the chiral
positive and negative sectors. Note that
$\sqrt\lambda\epsilon(\lambda)$ is the relative error in the
determination of the inverse square root, and all the integrals depend
only on this relative error.  Since there is no reason for $\rho(0)$ to
vanish, it is clear that the error in the expansion must remain under
control even as $\lambda\to0$.  Clearly, this is impossible to arrange
in polynomial expansions for $1/\sqrt\lambda$. However, a
finite sample of gauge configurations does not need full control over
$\epsilon(0)$, but only for $\epsilon(\lambda_<)$, where $\lambda_<$ is
the smallest eigenvalue encountered in the sample. To achieve this
while optimizing CPU costs on configurations where all the eigenvalues
are much larger requires the algorithm to be adaptive.

It was assumed above that no deflation has been performed, or that
deflation has been performed with no arithmetic errors. We comment on
the effects of deflation in Section \ref{sc.comp}.

\subsection{Rational approximation}
  In case of a rational function approximation to $M^{-1/2}$, one writes 
the operator $L$ as
\beq
   L=\sum_{i=1}^\nor\frac{b_i}{M+d_i}
\eeq

$E$ here depends on the order $\nor$ and the accuracy of the inversion
of $(M + d_i)$.  If the inversion can be achieved with infinite precision, 
then $L$ is a matrix again which commutes with $M$ and the analysis 
of the previous subsection applies in full. If, on the other hand, the error 
due to the inversion dominates, then for many algorithms, such as the
Conjugate Gradient, $L$ depends explicitly on the
vector $\Phi$ in a complicated way and it is not a matrix. One has to compute 
the different errors explicitly and study their behavior as in iterative 
methods.  Thus the behavior of errors from rational approximation case 
interpolates between that of the polynomial approximation and an iterative 
method according to the relation between the order and the precision of
the inversion.

\section{Fixed order: Chebychev approximation}\label{sc.ca}

The first use of the polynomial approximation utilized Legendre
polynomials \cite{hepilu}. Later the same group proposed a more robust
version using the Chebychev approximation (CA) \cite{hepile}.  As is well
known, when expanding any function, $f(z)$ in a fixed range $z_{\min}\le
z \le z_{\max}$, to a given order $\nor$ through orthogonal polynomials,
the use of Chebychev polynomials minimizes the maximum error on the
function to be approximated.

\begin{figure}[htb]
   \begin{center}\scalebox{0.5}{\includegraphics{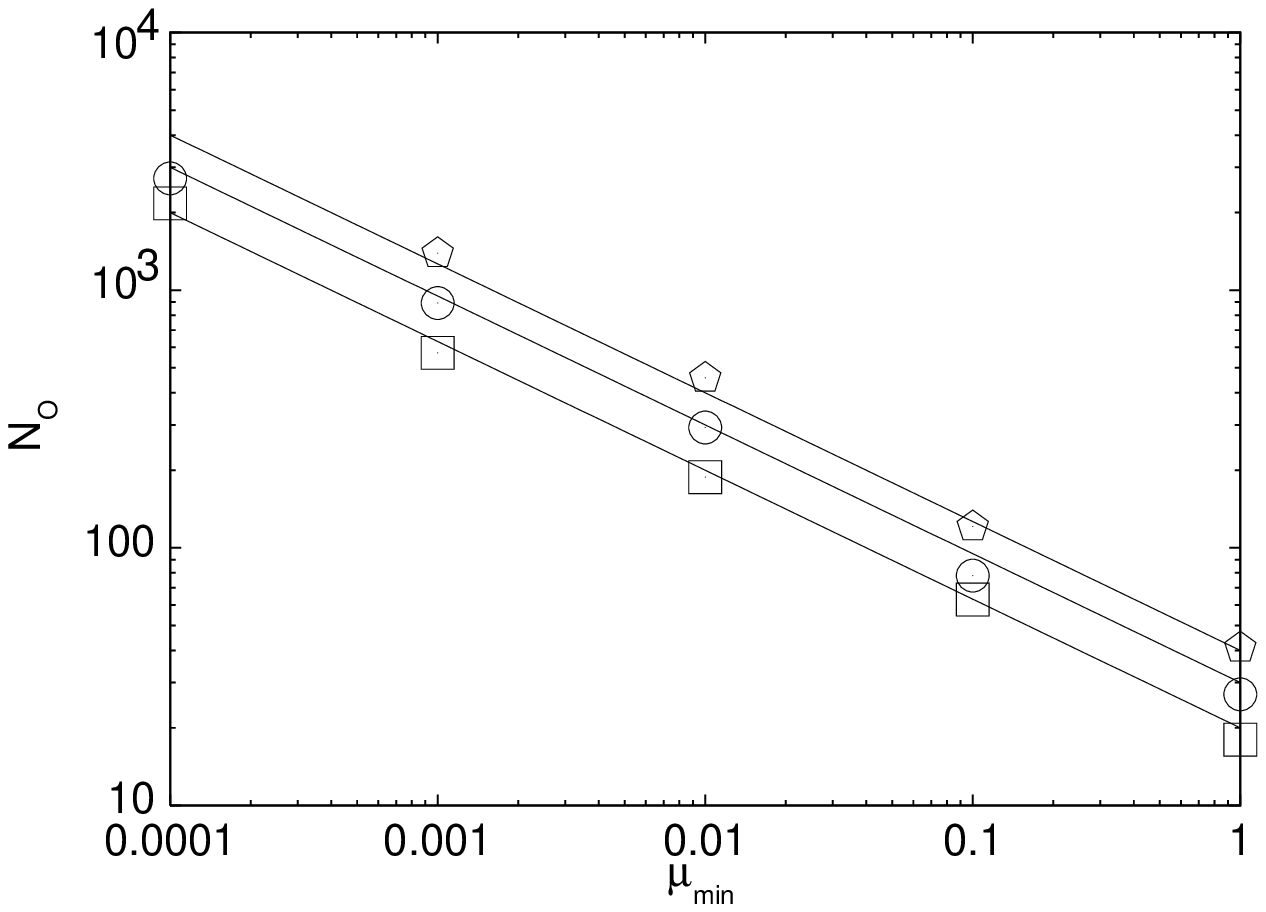}}
   \scalebox{0.5}{\includegraphics{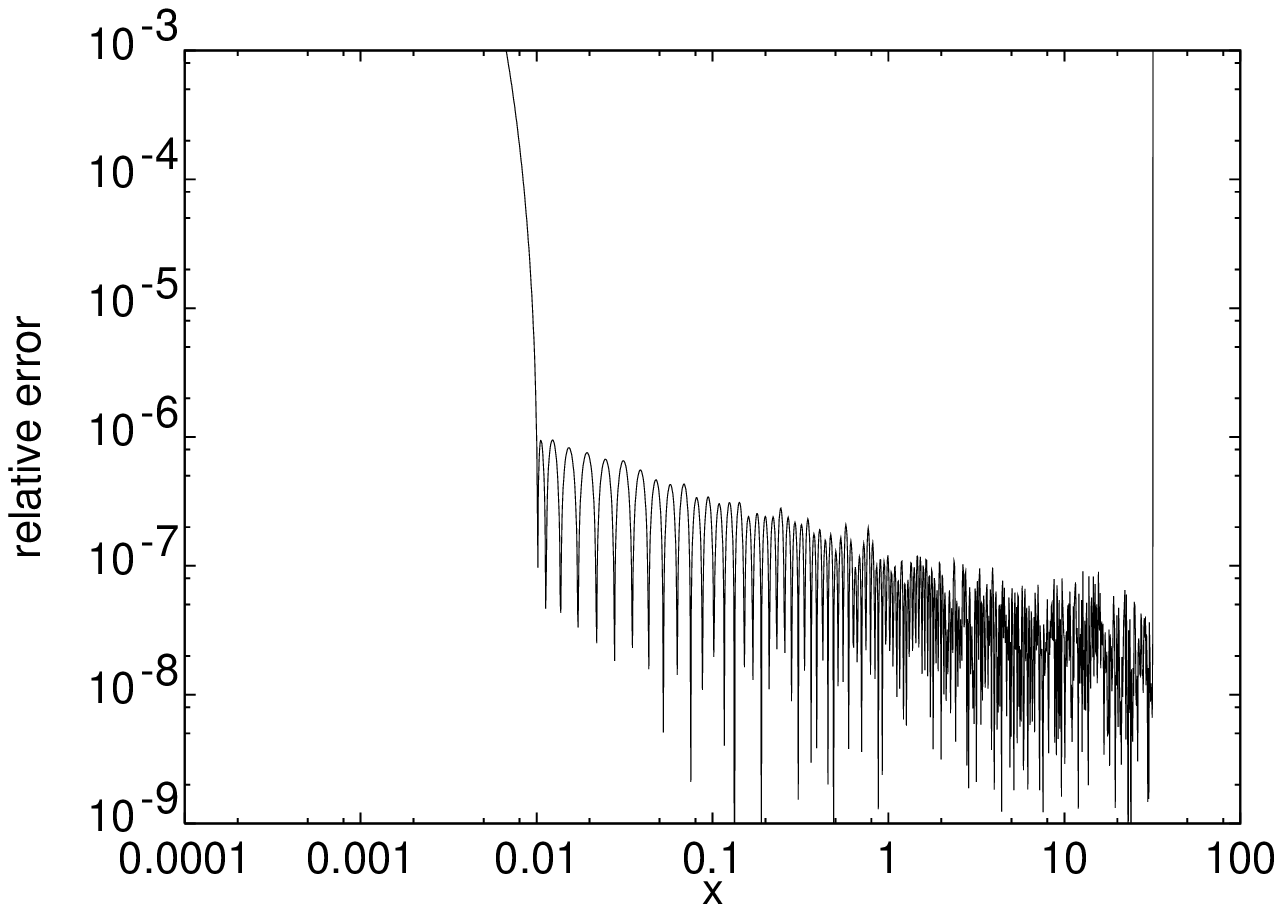}}\end{center}
   \caption{The panel on the left shows the order of Chebychev expansion,
      $\nor$ required to reach an accuracy of $10^{-3}$ (boxes),
      $10^{-4}$ (circles) and $10^{-6}$ (pentagons) in the range
      $[\mu_{\min},32]$ under variation of $\mu_{\min}$ (the lines are
      proportional to $\log(1/\varepsilon)/\sqrt \mu_{\min}$). The panel on
      the right shows the error for a Chebychev expansion in the range
      $[0.01,32]$ with $\nor=400$.}
\label{fg.chacc}\end{figure}

\begin{table}[hbtp]
  \begin{center}\begin{tabular}{|r||c|c|c||c|} \hline
    $\nor$ & $\ehf$ & $\egw$ & $\ecc$ & time \\  \hline
      30&$ 0.273\E{1}$&$ 0.273\E{1}$&$ 0.273\E{1}$& 0.3\\
      65&$ 0.174\E{2}$&$ 0.174\E{2}$&$ 0.174\E{2}$& 0.7\\
     100&$ 0.136\E{3}$&$ 0.136\E{3}$&$ 0.136\E{3}$& 1.1\\
     135&$ 0.113\E{4}$&$ 0.113\E{4}$&$ 0.113\E{4}$& 1.5\\
     165&$ 0.140\E{5}$&$ 0.140\E{5}$&$ 0.140\E{5}$& 1.8\\
     200&$ 0.125\E{6}$&$ 0.125\E{6}$&$ 0.125\E{6}$& 2.2\\
     235&$ 0.113\E{7}$&$ 0.113\E{7}$&$ 0.113\E{7}$& 2.5\\
     270&$ 0.102\E{8}$&$ 0.102\E{8}$&$ 0.102\E{8}$& 2.9\\
     300&$ 0.134\E{9}$&$ 0.134\E{9}$&$ 0.134\E{9}$& 3.4\\
     330&$ 0.174\E{10}$&$ 0.174\E{10}$&$ 0.174\E{10}$& 3.7\\
     360&$ 0.230\E{11}$&$ 0.230\E{11}$&$ 0.230\E{11}$& 4.1\\
     390&$ 0.391\E{12}$&$ 0.391\E{12}$&$ 0.391\E{12}$& 4.4\\
     420&$ 0.210\E{12}$&$ 0.209\E{12}$&$ 0.209\E{12}$& 4.8\\
     450&$ 0.226\E{12}$&$ 0.226\E{12}$&$ 0.226\E{12}$& 4.9\\ \hline
  \end{tabular}\end{center}
  \caption{Runs with the CA adjusted to the interval $[0.032,32]$ for
     varying $\nor$ on the configuration A. The last column gives the
     CPU seconds used on a \machine.}
\label{tb.cheby1}\end{table}

\begin{table}[hbtp]
  \begin{center}\begin{tabular}{|r||c|c|c||c|} \hline
    $\nor$ & $\ehf$ & $\egw$ & $\ecc$ & time \\  \hline
     100&$0.236\E{1}$&$0.236\E{1}$&$0.236\E{1}$& 1.1\\
     500&$0.294\E{2}$&$0.294\E{2}$&$0.294\E{2}$& 5.4\\
    1000&$0.485\E{3}$&$0.485\E{3}$&$0.485\E{3}$&10.8\\
    1500&$0.108\E{3}$&$0.108\E{3}$&$0.108\E{3}$&16.9\\
    2000&$0.292\E{4}$&$0.292\E{4}$&$0.292\E{4}$&21.8\\
    3000&$0.254\E{5}$&$0.254\E{5}$&$0.254\E{5}$&32.7\\
    4000&$0.267\E{6}$&$0.267\E{6}$&$0.267\E{6}$&43.7\\ \hline
  \end{tabular}\end{center}
  \caption{Runs with the CA adjusted to the interval $[7.2\E{5},32]$ on
     the configuration B. The last column shows the CPU seconds used
     on a \machine.}
\label{tb.cheby2}\end{table}

\begin{table}[hbtp]
  \begin{center}\begin{tabular}{|r||c|c|c||c|} \hline
    $\nor$ & $\ehf$ & $\egw$ & $\ecc$ & time \\  \hline
     100 &$0.253\E1$&$0.253\E1$&$0.253\E1$& 1.1\\
     500 &$0.788\E2$&$0.788\E2$&$0.788\E2$& 5.5\\
    1000 &$0.471\E2$&$0.471\E2$&$0.471\E2$&10.9\\
    1500 &$0.395\E2$&$0.395\E2$&$0.395\E2$&16.2\\
    3000 &$0.460\E2$&$0.460\E2$&$0.460\E2$&32.4\\ \hline
  \end{tabular}\end{center}
  \caption{Runs with the CA adjusted to the interval $[8.9\E9,32]$ on
     the configuration C. The last column shows the CPU seconds used
     on a \machine.}
\label{tb.cheby3}\end{table}

For the function $1/\sqrt z$, the coefficients of the Chebychev expansion
for $z_{\min}\le z\le z_{\max}$ to order $\nor$ are
\beq
   C_k = \frac{2\sqrt{2}}{\nor}\sum_{j=1}^\nor
      \frac{\cos((k-1)(j- \frac{1}{2})\pi/\nor)}{[z_{\min}+z_{\max}
		+(z_{\min}-z_{\max})\cos((j-\frac{1}{2})\pi/\nor)]^{1/2}}.
\label{coefcheby}\eeq
Applying this approximation to a matrix $M$ corresponds to finding
$L\Phi$ (see Eq.\ \ref{approx}) by the expansion---
\beq
   L\Phi = \frac{1}{2}c_1\Phi^{(0)} +\sum_{k=2}^\nor c_k \Phi^{(k-1)},
\label{cheby}\eeq
where the successive vectors $\Phi^{(n)}$ can be found by the iteration
$\Phi^{(0)}=\Phi$ and
\beq
   \Phi^{(n)} = \frac2{z_{\max}-z_{\min}}\left[2 M \Phi^{(n-1)} 
	- (z_{\max}+z_{\min}) \Phi^{(n-1)}\right] - \Phi^{(n-2)},
\label{itcheby}\eeq
for $n\ge2$. For $n=1$ the term $\Phi^{(-1)}$ is dropped from the
recursion.

There are various sources of error, which we now analyze in
succession. For the scalar version of the algorithm (or for each
eigenvector of $M$ separately), with fixed
$z_{\min}$, $z_{\max}$ and parameter $\nor$, there is an error in the
approximation of $1/\sqrt z$ which we call $\epsilon(z)$. For
computations at arbitrary precision with a fixed range,
$[z_{\min},z_{\max}]$, $\epsilon(z)$ depends entirely on $\nor$.
To keep the absolute relative error bounded, $|\epsilon(z)|\sqrt 
z\le\varepsilon$, as $z_{\min}$ changes, we must tune
\beq
   \nor\propto\log\left(\frac1\varepsilon\right)\sqrt\frac1{z_{\min}}\simeq
       \log\left(\frac1\varepsilon\right)\sqrt\kappa,
\label{norsc}\eeq
as shown in Figure \ref{fg.chacc}. The last expression follows if we choose
$z_{\min}=\lambda_{\min}$ and $z_{\max}=\lambda_{\max}$.

However, as shown in the right panel of the figure, tuning $\nor$ in 
this manner causes the relative error outside the range to
blow up. The CA has no tolerance to violations of the requirement on
the range. This is easy to understand. In any polynomial approximation,
with decreasing $z_{\min}$ larger values of $\nor$ are required for
keeping the error fixed within the interval $[z_{\min},z_{\max}]$.
However, outside this interval, the error then increases as
\beq
   \epsilon(z)\propto\left(\frac{2z-z_{\min}-z_{\max}}{z_{\max}
					      -z_{\min}}\right)^\nor,
   \qquad\qquad{\rm for\ } z<z_{\min}{\rm\ or\ }z>z_{\max},
\label{error}\eeq
and hence the error increases faster as $z_{\min}$ decreases. Solving
this for $z$, given some fixed value of $\epsilon(z)/\varepsilon$, we can easily
see that for large $\kappa$ and fixed precision $\varepsilon$, 
\beq
   {\cal A}\propto \exp(-\alpha\sqrt\kappa)
\label{adaptca}\eeq
where $\alpha$ is some number.

The effect of finite arithmetic precision can also be analyzed easily
since the iteration in Eq.\ (\ref{itcheby}) is linear.  Any arithmetic
error, $\delta^{(m)}$, in $\Phi^{(m)}$ of the order of the machine
precision remains in control whenever all eigenvalues, $\lambda$ of $M$
satisfy $z_{\min} \le \lambda \le z_{\max}$. If any eigenvalue of $M$
lies outside this range, then the iteration magnifies the error
geometrically---
\beq
   \delta^{(m+n)} \simeq 
   \left(\frac{2\lambda_{\min}-z_{\min}-z_{\max}}{z_{\max}
	   -z_{\min}}\right)^n\delta^{(m)},
\label{deflerr}\eeq
This is the vector version of the low adaptability of this algorithm.
If estimates of $\lambda_{\min}$ and $\lambda_{\max}$ for $M$ are
available, then, in view of this instability, it is best to choose
$z_{\min}<\lambda_{\min}$ and $z_{\max}>\lambda_{\max}$.

The complexity of this algorithm is clearly dominated by the time required
to operate upon a vector by the matrix $M$ in the iterations in Eq.\
(\ref{itcheby}). For the Wilson-Dirac matrix
this time is of order $V$. Neglecting the time taken for scalar operations
in the remainder of the algorithm, and also the order $V$ time for vector
additions, in comparison with this, we have---
\beq
   {\cal C}_{\scriptscriptstyle CA} \simeq w\nor V\simeq
   w'V\log\left(\frac1\varepsilon\right)\sqrt\kappa,
\label{compca}\eeq
where $wV$ is the complexity of operating upon a vector by $M$ and $w,w'$
are constants.  Apart from the gauge configuration, in QCD applications
the storage required is for the three vectors needed
for the iteration in Eq.\ (\ref{itcheby}). The
space complexity is therefore
\beq
   {\cal S}_{\scriptscriptstyle CA} = 8N_c(N_c+3)V,
\label{spacca}\eeq
(for $N_c$ colors) neglecting storage for scalars.

With a fixed $\nor$, the precision of the algorithm deteriorates sharply
when one or more of the eigenvalues of $M$ lie outside the interval
$[z_{\min},z_{\max}]$, as shown in Figure \ref{fg.chacc} and by the low
value of $\cal A$ in Eq.\ (\ref{adaptca}). This is often sought to be
corrected for by deflating, \ie, explicitly dealing with the eigenspace
of the lowest eigenmodes, and applying the algorithm to the orthogonal
space. This would keep the accuracy constant as the condition number
changes. If $\ndf$ vectors need to be deflated, then the contribution
to $\cal C$ clearly increases as $(\ndf V)^2$, since each vector has to be
orthogonalized with respect to every other. Also, $\cal S$ increases as
$\ndf V$ due to the necessity of storing the vectors. In order to achieve
a target precision, $\varepsilon$, on all gauge configurations, we are
forced to deflate all vectors with $\lambda<z_{\min}$. Working with a
fixed $\ndf$ forces us to do unnecessarily large amount of work on most
configurations, while still failing on a small set of configurations.
As a result, $\ndf$ has to be chosen appropriately for each configuration.
With deflation then we have
\beqa
\nonumber
   {\cal C}_{CA} &=& w'V\log\left(\frac1\varepsilon\right)\sqrt{\kappa_{eff}}
	  +w''V^2\langle\ndf^2\rangle,\\
   {\cal S}_{CA} &=& 8N_c(N_c+3)V+8N_c\langle\ndf\rangle V,
\label{complxca}\eeqa
where the angular brackets denote averages over gauge configurations,
$w''$ is a constant independent of $V$ and $\ndf$,
and $\kappa_{eff}$ is the condition number of the matrix after
deflating $\ndf$ vectors. With careful programming we can arrange to
make $w''<w'$, although they cannot differ by orders of magnitude
($w''$ can depend on $\varepsilon$ and $\lambda_{\min}$).

We have not investigated the expectation values of $\ndf$. However, 
when we change the volume at fixed physics, we
expect that $\langle\ndf\rangle\propto V\overline\rho$, where
$\overline\rho$ is the average density of eigenvalues of $M$ near
$\lambda_{\min}$.  Since 
deflation is designed to hold $\kappa_{eff}$ fixed, this means that for
large enough $V$ the complexity ${\cal C}_{CA}\propto V^4$. On the
other hand, $\kappa$ should generically grow linearly in $V$. Hence, on
sufficiently large volumes, without deflation we would have ${\cal
C}_{CA}\propto V^{3/2}$. For best actual performance, one would have to
tune the playoff between these two limits.

A further complication arises in CA, and indeed, in any method which
utilizes a polynomial expansion. For any fixed finite precision, the
deflation of $\Phi^{(0)}$ is inaccurate since each component of the
deflated vector is in error at least in the least significant bit.
Since the CA iteration of Eq.\ (\ref{itcheby}) is not stable on the
deflated eigenspace, this error blows up geometrically as in
Eq.\ (\ref{deflerr}). Consequently, more and more bits are corrupted, 
as the iteration proceeds, and the process may eventually render the
whole computation unusable.  Note that this problem becomes more acute
with decreasing $\kappa$, even if several eigenvectors are deflated.
To prevent the error from swamping the result in this fashion, one has
to reorthogonalize $\Phi^{(n)}$ repeatedly (this process itself is not
free of complications, see \cite{golub}). This involves finding $\ndf$
dot products and subtracting $\ndf$ vectors. While this is crucial in
maintaining the accuracy of the result, it does not change the
complexity, and we still have the results in Eq.\ (\ref{complxca}).

A different approach, and the one we have adopted for our tests, is
to start the algorithm by making an estimate of $\lambda_{\min}$ and
$\lambda_{\max}$ and then to select $\nor$ accordingly. This obviates
any need for deflation, and controlling the rounding errors in such a
method. The algorithm is well-behaved, both with respect to precision and
propagation of rounding errors, whenever $z_{\min} \le \lambda_{\min}
\le \lambda_{\max} \le z_{\max}$. However, in this case the complexity
rises as $\sqrt\kappa$.

As shown in Table \ref{tb.cheby1}, the algorithm performs well on
configuration A. $\nor$ needed to achieve a given value of $\ehf$ is
seen to rise logarithmically, as argued above.  Consequently, $\ehf$
can be made very small and the required chiral properties obtained at
any desired precision. Note also that the equalities (\ref{rel1})
are exactly satisfied, as expected, and remain so for 
configurations B and C too, as seen from the Tables \ref{tb.cheby2} 
and \ref{tb.cheby3} respectively.  However, the algorithm
was found to be extremely slow for these configurations and saturated
at $\nor \sim 1500$ with $\ehf=4\times 10^{-3}$ on the configuration C,
leading to the same precision for both the GW relation and the
unit circle property of $D$.

\section{Fixed order: Optimized Rational Approximation}\label{sc.ora}

\begin{figure}[htb]
   \begin{center}\includegraphics{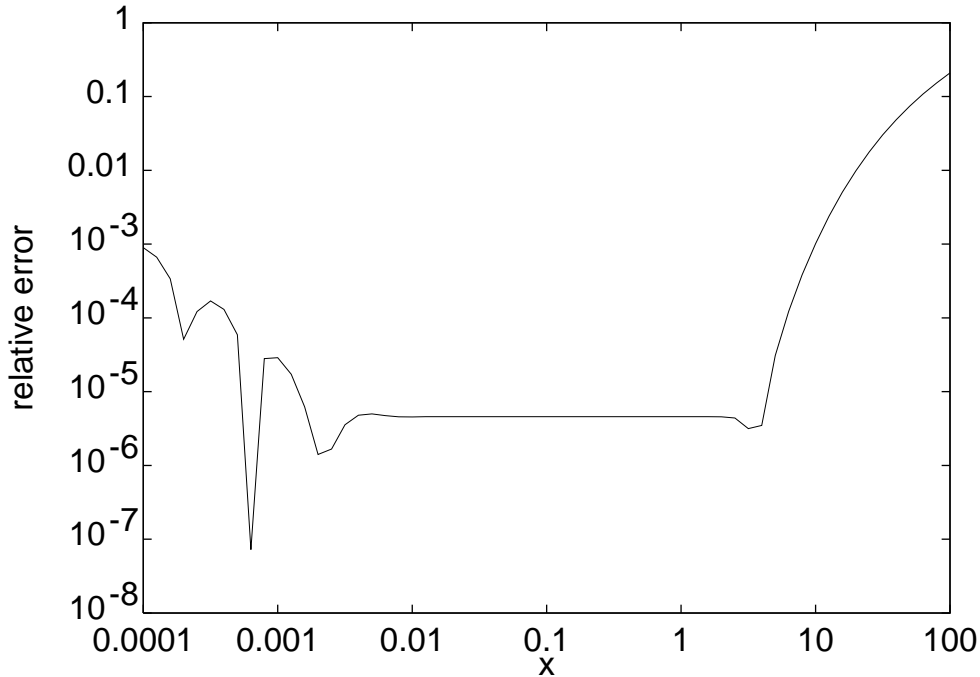}\end{center}
   \caption{The relative error, $|\epsilon(z)|\sqrt z$, in the ORA with
      $\nor=14$ fitted to the range $[0.01,1]$ for computation in
      and outside the range.}
\label{fg.oraacc}\end{figure}

\begin{table}[hbtp]\begin{center}
\begin{tabular}{|c||c|c||c|c||c|c|}
\hline
$\alpha$ & $10^3\ehf$ & $10^3\egw$ & $10^3\ehf$ & $10^3\egw$ & $10^3\ehf$ & $10^3\egw$ \\
\hline
1.00 & 11.2  & 11.3 & 11.2   & 11.2  & 11.2   & 11.2 \\
0.50 & 1.74  & 2.01 & 1.44   & 1.44  & 1.44   & 1.44 \\
0.20 & 0.987 & 1.40 & 0.0780 & 0.116 & 0.0257 & 0.0273 \\
0.15 & 0.986 & 1.40 & 0.0744 & 0.114 & 0.0112 & 0.0143 \\
0.12 & 0.986 & 1.40 & 0.0744 & 0.114 & 0.0110 & 0.0142 \\
0.10 & 0.986 & 1.40 & 0.0744 & 0.114 & 0.0111 & 0.0142 \\
0.08 & 0.986 & 1.40 & 0.0744 & 0.114 & 0.0111 & 0.0143 \\
0.05 & 0.986 & 1.40 & 0.0744 & 0.114 & 0.0111 & 0.0142 \\
\hline
\end{tabular}\end{center}
  \caption{Tuning $\alpha$ in ORA for a fixed configuration with three
     different values of $\ecg$. Since this fixes the upper part of the
     range of eigenvalues, we expect little change from one
     configuration to another, and $\alpha=0.1$ is a global choice.}
\label{tb.oratuning}\end{table}

The first algorithm to compute $M^{-1/2}$ through a rational expansion was
a polar formula introduced by Neuberger \cite{neu}.
An improved version \cite{nara}, called the optimized rational
approximation, uses coefficients obtained numerically
to give the approximation
\beq
  L\Phi=\sqrt\alpha\,\left[c_0
      + \sum_{k=1}^\nor\frac{c_k}{\alpha M+d_k}\right]\Phi,
\label{ora}\eeq
where $\alpha$ is an arbitrary scale whose
choice we describe later, and the values of $c_k$ and $d_k$ for
a given $\nor$ are obtained by optimizing the fit on some  fixed interval 
$[z_{\min},z_{\max}]$ using the Remez algorithm (see \cite{nara} for
details).
The inversion of $(\alpha M+d_k)$ is made by a
multimass conjugate gradient stopped according to a value $\ecg$ for
the residual.  This version is used in particular in \cite{nara2},
\cite{liu}, and \cite{degrand}.

\begin{table}[hbtp]
  \begin{center}\begin{tabular}{|c|r||c|c|c||c|} \hline
    $\ecg$ & $\ncg$ & $\ehf$ & $\egw$ & $\ecc$ & time \\ \hline
    $\e{1}$& 25&$  0.179\E{1}$&$  0.257\E{1}$&$  0.539\E{1}$& 0.5\\
    $\e{2}$& 63&$  0.986\E{3}$&$  0.140\E{2}$&$  0.614\E{2}$& 1.1\\
    $\e{3}$& 99&$  0.744\E{4}$&$  0.114\E{3}$&$  0.507\E{3}$& 1.6\\
    $\e{4}$&134&$  0.111\E{4}$&$  0.142\E{4}$&$  0.397\E{4}$& 2.1\\
    $\e{5}$&166&$  0.916\E{5}$&$  0.919\E{5}$&$  0.966\E{5}$& 2.6\\
    $\e{6}$&198&$  0.914\E{5}$&$  0.914\E{5}$&$  0.914\E{5}$& 3.1\\
    \hline
  \end{tabular}\end{center}
  \caption{Runs with the ORA for $\nor=14$ optimized in the interval
     $[0.01,1]$ and $\alpha=0.1$ on configuration A.
     The last column gives the CPU seconds used on a \machine.}
\label{tb.ora1}\end{table}

\begin{table}[hbtp]
  \begin{center}\begin{tabular}{|c|r||c|c|c||c|} \hline
    $\ecg$ & $\ncg$ & $\ehf$ & $\egw$ & $\ecc$ & time \\ \hline
   $\e{1}$& 71&$0.148\E{1}$&$0.243\E{1}$&$0.542\E{1}$&1.2\\
   $\e{2}$&332&$0.767\E{4}$&$0.113\E{3}$&$0.799\E{2}$&5.0\\
   $\e{3}$&363&$0.220\E{4}$&$0.229\E{4}$&$0.499\E{2}$&5.5\\
   $\e{4}$&395&$0.208\E{4}$&$0.208\E{4}$&$0.160\E{2}$&6.0\\
   $\e{5}$&426&$0.208\E{4}$&$0.208\E{4}$&$0.209\E{4}$&6.4\\
    \hline
  \end{tabular}\end{center}
  \caption{Runs with the ORA for $\nor=14$ optimized in the interval
     $[0.01,1]$ and $\alpha=0.1$ on configuration B.
     The last column shows the CPU seconds used on a \machine.}
\label{tb.ora2}\end{table}

\begin{table}[hbtp]
  \begin{center}\begin{tabular}{|c|r||c|c|c||c|} \hline
    $\ecg$ & $\ncg$ & $\ehf$ & $\egw$ & $\ecc$ & time \\ \hline
    $\e1$& 317 &$0.161\E1$&$0.246\E1$&$0.547\E1$&  4.9\\
    $\e2$& 781 &$0.333\E2$&$0.333\E2$&$0.111\E1$& 11.9\\
    $\e3$& 815 &$0.333\E2$&$0.333\E2$&$0.771\E2$& 12.5\\
    \hline
  \end{tabular}\end{center}
  \caption{Runs with the ORA for $\nor=14$ optimized in the interval
     $[0.01,1]$ and $\alpha=0.1$ on configuration C.
     The last column shows the CPU seconds used on a \machine.}
\label{tb.ora3}\end{table}

Taking the values of $c_k$ and $d_k$ for $\nor=14$, $z_{\min}=0.01$ and
$z_{\max}= 1$ from \cite{nara}, we show in Figure \ref{fg.oraacc} the relative
accuracy, $|\epsilon(z)|\sqrt z$ for the expansion both inside and outside the
fitted range. It is clear from the figure that the algorithm rapidly degrades
outside the chosen range. A numerical computation shows that 
\beq
   {\cal A} \approx 1,
\label{adaptora}\eeq
so that it has much higher adaptability than the CA. Nevertheless, it
makes large errors on the eigenvalues of $M$
which are greater than $z_{\max}$. It is easy to see that many
eigenvalues are greater than unity, and a scaling factor $\alpha$ is
therefore needed to bring these into range. The tuning of $\alpha$ is
shown in Table \ref{tb.oratuning}.  For the conjugate gradient
inversion of each term, when $\ecg$ is large, it determines the
accuracy of the solution. Hence $\ecg$ must be kept small enough so that
the accuracy is $\epsilon$.

In order to specify the scaling of the CPU time in each algorithm
with various parameters of the problem, we count the complexity
of the method. For this algorithm, $\cal C$ is proportional to the
number of steps of the Conjugate Gradient inversion, $\ncg$. It can be
proved that $\ncg$ grows no faster than $\sqrt{\kappa}$.  However,
from the observed
convergence rate of the CG iterations (shown in Section \ref{sc.comp}),
we see that $\ncg\propto \log(1/\ecg)\log\kappa$. This expression can
be used when $\ecg$ is tuned to be smaller than the error shown in
Figure \ref{fg.oraacc}.  For each step of the multimass CG inversion,
the complexity is dominated by the time required to operate upon a
vector by the matrix $M$. For the Wilson-Dirac matrix this is of order
$V$. Neglecting the time taken for scalar operations and also the order
$V$ time for vector additions, we have---
\beq
   {\cal C}_{ORA} \simeq w\ncg V \simeq w'V\log\left(\frac1\ecg\right)\log\kappa,
\label{compora}\eeq
where $wV$ is the complexity of operating upon a vector by $M$, and
$w'$ is a constant. The dependence of $N_{CG}$ on $V$ is very weak for 
realistic $\ncg$ and is neglected. The memory requirement is essentially for the
storage of the gauge configuration and for $2+2\nor$ vectors required
to build up the approximation. The space complexity is therefore
\beq
   {\cal S}_{ORA} = 8N_c(N_c+2+2\nor)V,
\label{spacora}\eeq
(for $N_c$ colors) neglecting storage for scalars.

With a fixed value of $\nor$, the precision of the algorithm
deteriorates when the eigenvalues of $M$ lie outside the range
$[z_{\min},z_{\max}]$, as shown in Figure \ref{fg.oraacc}.  For
$D_w$, the highest eigenvalue remains in the vicinity of 32 for most
configurations, whereas the lowest eigenvalue may fluctuate by several
orders of magnitude. The large eigenvalues are brought into range by
tuning $\alpha<1$ as shown already. However, this drives the lowest
eigenvalues further out of the range, thus degrading performance. As a
result, it is necessary to deflate, {\sl i.e.\/}, explicitly deal with
the eigenspace of the lowest eigenmodes, and apply ORA on the
orthogonal space in order to keep a constant accuracy as the condition
number changes. As before, this changes the complexity to
\beqa
\nonumber 
   {\cal C}_{ORA} &=& w'V\log\left(\frac1\ecg\right)\log\kappa_{eff}
	 + w''\langle\ndf^2\rangle V^2,\\ 
   {\cal S}_{ORA} &=& 8N_c(N_c+2+2\nor)V + 8N_c\langle\ndf\rangle V.
\label{oradefl}\eeqa 
where the angular brackets denote averaging over the sample of
configurations used. From available data on the growth of $\nor$
required to keep the relative error fixed with growing $\kappa$
\cite{nara} it seems that it is better to increase $\nor$ rather than
to keep $\kappa_{eff}$ fixed by increasing $\langle\ndf^2\rangle$ with
increasing volume at fixed physics.

In Tables \ref{tb.ora1}, \ref{tb.ora2} and \ref{tb.ora3} we show the
results of our numerical tests of the ORA using the set of $c_k$ and $d_k$ 
for $\nor=14$ from \cite{nara} for the three configurations already described. 
One sees that with higher precision of inversion, $\epsilon_{CG}$, the
relations (\ref{rel1}) indeed get satisfied well. 
The gradual deterioration of the performance of ORA with
fixed $\nor$ in going to larger $\kappa$ is clear from the tables. This
indicates that the performance of ORA may improve if a degree of
adaptability can be built into the algorithm by, for example, allowing
for changes in $\nor$ because with $\nor=14$, we cannot obtain
better precision than 0.00001, 0.00002 and 0.003 for the GW relation for
configurations A, B and C respectively. 

\section{Fixed order: Zolotarev approximation}\label{sc.za}

\begin{figure}[htb]
   \begin{center}\includegraphics{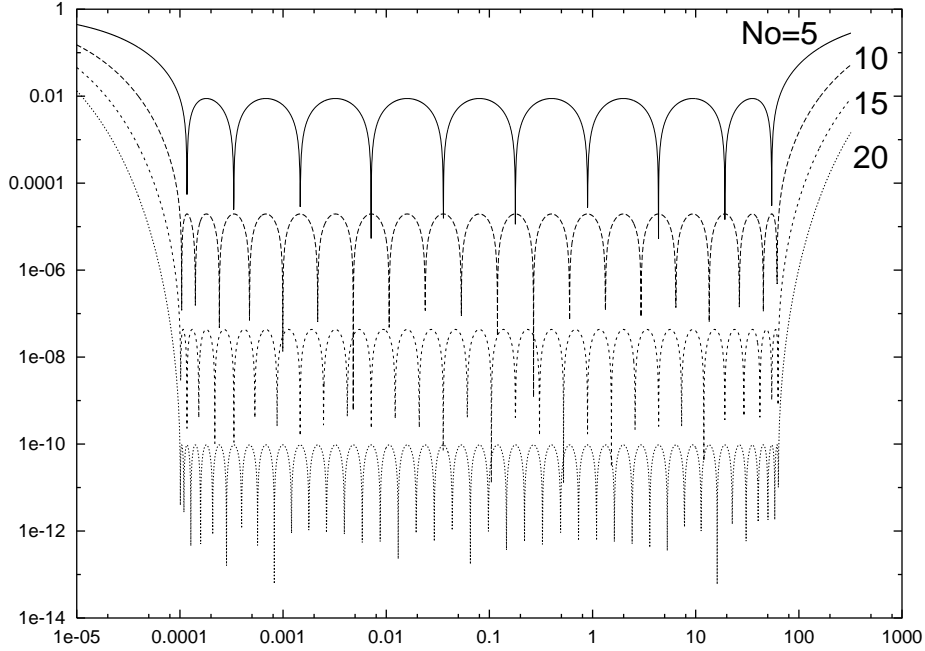}\end{center}
   \caption{The relative error, $|\epsilon(z)|\sqrt z$  for ZA in the 
      range $[10^{-4},32]$ for computation in and outside the range
      for various $\nor$.}
\label{fg.zaacc}\end{figure}

\begin{table}[hbtp]
\begin{center}\begin{tabular}{|r|c|r||c|c|c||c|} \hline
  $\nor$ & $\ecg$ & $\ncg$ & $\ehf$ & $\egw$ & $\ecc$ & time \\ \hline
    5&$\e{1}$&  22&$ 0.207\E{1}$&$ 0.281\E{1}$&$ 0.534\E{1}$& 0.3\\
    6&$\e{2}$&  56&$ 0.144\E{2}$&$ 0.216\E{2}$&$ 0.612\E{2}$& 0.7\\
    7&$\e{3}$&  91&$ 0.114\E{3}$&$ 0.160\E{3}$&$ 0.668\E{3}$& 1.2\\
    8&$\e{4}$& 126&$ 0.984\E{5}$&$ 0.150\E{4}$&$ 0.526\E{4}$& 1.7\\
    9&$\e{5}$& 159&$ 0.849\E{6}$&$ 0.123\E{5}$&$ 0.467\E{5}$& 2.2\\
   10&$\e{6}$& 191&$ 0.808\E{7}$&$ 0.117\E{6}$&$ 0.365\E{6}$& 2.6\\
   11&$\e{7}$& 223&$ 0.808\E{8}$&$ 0.119\E{7}$&$ 0.343\E{7}$& 3.1\\
   13&$\e{8}$& 259&$ 0.616\E{9}$&$ 0.924\E{9}$&$ 0.288\E{8}$& 3.8\\
   14&$\e{9}$& 295&$0.566\E{10}$&$0.798\E{10}$&$ 0.289\E{9}$& 4.4\\
   15&$\e{10}$&326&$0.557\E{11}$&$0.793\E{11}$&$0.296\E{10}$& 5.0\\
   16&$\e{11}$&357&$0.553\E{12}$&$0.723\E{12}$&$0.225\E{11}$& 5.5\\
   \hline
\end{tabular}\end{center}
  \caption{Runs with the ZA in the interval $[0.035,32]$ for
     configuration A. The last column gives the CPU seconds
     used on a \machine.}
\label{tb.zolo1}\end{table}

\begin{table}[hbtp]
\begin{center}\begin{tabular}{|r|c|r||c|c|c||c|} \hline
  $\nor$ & $\ecg$ & $\ncg$ & $\ehf$ & $\egw$ & $\ecc$ & time \\ \hline
     7&$\e{1}$& 71&$ 0.148\E{1}$&$0.243\E{1}$&$0.540\E{1}$& 0.9\\
    10&$\e{2}$&331&$ 0.876\E{4}$&$0.130\E{3}$&$0.798\E{2}$& 4.5\\
    12&$\e{3}$&362&$ 0.833\E{5}$&$0.110\E{4}$&$0.499\E{2}$& 5.2\\
    14&$\e{4}$&394&$ 0.814\E{6}$&$0.101\E{5}$&$0.160\E{2}$& 5.9\\
    16&$\e{5}$&426&$ 0.741\E{7}$&$0.902\E{7}$&$0.366\E{6}$& 6.6\\
    18&$\e{6}$&457&$ 0.744\E{8}$&$0.976\E{8}$&$0.297\E{7}$& 7.3\\
    20&$\e{7}$&488&$ 0.692\E{9}$&$0.103\E{8}$&$0.313\E{8}$& 8.1\\
    22&$\e{8}$&516&$0.693\E{10}$&$0.963\E{10}$&$0.278\E{9}$& 8.9\\
    24&$\e{9}$&544&$ 0.644\E{11}$&$0.903\E{11}$&$0.253\E{10}$& 9.7\\
    26&$\e{10}$&572&$ 0.715\E{12}$&$0.906\E{12}$&$0.244\E{11}$&10.4\\
   \hline
\end{tabular}\end{center}
  \caption{Runs with the ZA in the interval $[7.2\times10^{-5},32]$
     for configuration B. The last column indicates the CPU seconds
     used on a \machine.}
\label{tb.zolo2}\end{table}

\begin{table}[hbtp]
\begin{center}\begin{tabular}{|r|c|r||c|c|c||c|} \hline
  $\nor$ & $\ecg$ & $\ncg$ & $\ehf$ & $\egw$ & $\ecc$ & time \\ \hline
   10 &$\e1$& 325 &$0.163\E1$&$0.247\E1$&$0.545\E1$&  4.5\\
   16 &$\e2$& 871 &$0.242\E4$&$0.125\E2$&$0.111\E1$& 13.4\\
   20 &$\e3$& 899 &$0.121\E5$&$0.107\E5$&$0.771\E2$& 14.9\\
   23 &$\e4$& 933 &$0.119\E6$&$0.105\E6$&$0.325\E2$& 16.3\\
   26 &$\e5$& 968 &$0.126\E7$&$0.103\E7$&$0.325\E2$& 17.7\\
   30 &$\e6$& 997 &$0.462\E8$&$0.482\E9$&$0.749\E9$& 19.4\\
   36 &$\e7$&1024 &$0.521\E8$&$0.150\E{10}$&$0.450\E{10}$& 21.7\\
   \hline
\end{tabular}\end{center}
  \caption{Runs with the ZA in the interval $[8.9\times10^{-9},32]$
     for configuration C. The last column indicates the CPU seconds
     used on a \machine.}
\label{tb.zolo3}\end{table}

The Zolotarev algorithm was explored in \cite{wupp} and used in \cite{ch}.
It is a rational expansion defined by
\beq
   L\Phi = \sum_{l=1}^\nor \left(\frac{b_l}{M+d_l}\right)\Phi,
\label{za}\eeq
in the range $[z_{\min},z_{\max}]$ (the smallest and largest
eigenvalues of $M$ must satisfy the conditions $z_{\min}\le
\lambda_{\min} \le \lambda_{\max} \le z_{\max}$), and the expansion
coefficients are
\beq
   d_l = c_{2l-1}\qquad{\rm and}\qquad
   b_l = d_0\frac{\prod_{i=1}^{\nor-1}(c_{2i}-c_{2l-1})}{
		  \prod_{i=1, i\ne l}^{\nor-1}(c_{2i-1}-c_{2l-1})}.
\label{zac}\eeq
The $c_l$'s are
\beq
   c_l = \frac{\sn^2(lK/2\nor;\sqrt{1-z_{\min}/z_{\max}})}{\cn^2(lK/2\nor;\sqrt{1-z_{\min}/z_{\max}})},
\label{zacl}\eeq
where the values of the Jacobi
elliptic functions, $\sn(u,k)=\sin \eta$ and $\cn(u,k)=\cos \eta$, are defined by the integral
\beq
   u(\sin(\eta)) = \int_0^{\sin(\eta)}\frac{dt}{\sqrt{(1-t^2)(1-k^2t^2)}}.
\label{jacobi}\eeq
The constant in Eq.\ (\ref{zacl}), $K=u(1)$, is the complete elliptic
integral.  When sn is near 0 or 1, high precision in the expressions of the 
coefficients of the corresponding $c$'s is essential.
The constant $d_0$ in Eq. (\ref{zac}) can be expressed in term of elliptic theta function
\cite{chiu2}, or {\it equivalently}, fixed by the condition \cite{wupp}
\beq
   \min_z\sum_{i=1}^\nor\left(\frac{\sqrt{z}b_l}{z+d_l}\right)+
   \max_z\sum_{i=1}^\nor\left(\frac{\sqrt{z}b_l}{z+d_l}\right)=2.
\label{dzero}\eeq

As in the ORA, the multimass CG which is used to invert the terms in
Eq.\ (\ref{za}) should have a stopping criterion, $\ecg$.
One advantage of ZA over ORA is that the
quantities $d_l$ in Eq.\ (\ref{za}) are larger than those in Eq.\
(\ref{ora}). As a result, a multimass CG inverter can evaluate this
approximation somewhat faster. Another advantage, emphasized in
\cite{wupp} is that $\nor$ required for a certain accuracy is smaller
for ZA than for ORA. It was found that a relative accuracy of better
than 1 part in $10^5$ is obtained for the interval $[0.01,1]$ with
$\nor\approx 6$ in ZA, as compared to 14 in ORA. As shown in Figure
\ref{fg.zaacc}, the relative error for ZA in the range $[10^{-4},32]$
does not require significantly higher $\nor$ for similar control over
error. However, the low adaptability, ${\cal A}\simeq 0.01$, means
that the coefficients should be computed over a range appropriate to
the condition number of the matrix. The main effect of increasing the
range for a fixed $\nor$ is to change the coefficients $b_l$ and $d_l$
in such a way that the logarithmic range of $d_l$ increases. We found that a factor
100 decrease in $z_{\min}$ (for fixed $z_{\max}=32$) led to a factor
20--30 decrease in the ratio of the minimum and maximum values of $d_l$.

The complexity and spatial complexity of ZA are very similar to that of
ORA. The complexity is dominated by the matrix-vector multiplication in
the CG inversions, and the memory requirement is dominated by the
vectors in the multimass CG. Hence
\beqa
\nonumber
   {\cal C}_{ZA} &\simeq& w'V\log\left(\frac1\ecg\right)\log\kappa,\\
   {\cal S}_{ZA} &=& 8N_c(N_c+2+2\nor)V,
\label{compza}\eeqa
where $w'$ is some constant, $N_c$ is the number of colors and $V$ is
the lattice volume. Since the effect of deflation is also similar
to that in ORA, we do not repeat that discussion here.

The performance of the Zolotarev algorithm in numerical tests
is summarized in Tables \ref{tb.zolo1}, \ref{tb.zolo2} and \ref{tb.zolo3}.
One needs to tune two algorithmic parameters, $\nor$ and $\epsilon_{CG}$
for better efficiency. For a given value of $\epsilon_{CG}$, we increase 
$\nor$ until a saturation in the value of error is evident.
For $\nor\approx 6 - 8$, the performance is similar to that of the ORA. The
improvement with increasing $\nor$ as $\kappa$ increases further indicates
that the ZA and ORA should both improve if $\nor$  is
allowed to change algorithmically with configuration.
Such a method can be constructed from the results of \cite{wupp},
when $\nor$ is increased until the maximum relative error (see 
Figure \ref{fg.zaacc}) attains a fraction ($<$1/2) of the desired accuracy.  
This can be implemented at the initialization step from the knowledge of
the minimum and maximum of the relative error (defined in LHS of 
Eq.(\ref{dzero}) or from the elliptic theta functions).

\section{Adaptive algorithm: conjugate gradient approximation}\label{sc.cga}

\begin{table}[hbtp]
  \begin{center}\begin{tabular}{|c||r||c|c|c||c|}\hline
     $\ecg$ & $\ncg$ & $\ehf$ & $\egw$ &  $\ecc$ &time \\ \hline
     $\e{1}$&  22&$  0.230\E{1}$&$  0.410\E{1}$&$  0.860\E{1}$& 0.5\\
     $\e{2}$&  55&$  0.178\E{2}$&$  0.325\E{2}$&$  0.124\E{1}$& 1.2\\
     $\e{3}$&  90&$  0.145\E{3}$&$  0.279\E{3}$&$  0.195\E{2}$& 2.1\\
     $\e{4}$& 125&$  0.121\E{4}$&$  0.266\E{4}$&$  0.148\E{3}$& 2.9\\
     $\e{5}$& 158&$  0.103\E{5}$&$  0.219\E{5}$&$  0.134\E{4}$& 3.6\\
     $\e{6}$& 190&$  0.925\E{7}$&$  0.182\E{6}$&$  0.926\E{6}$& 4.2\\
     $\e{7}$& 222&$  0.899\E{8}$&$  0.172\E{7}$&$  0.839\E{7}$& 5.0\\
     $\e{8}$& 257&$  0.859\E{9}$&$  0.175\E{8}$&$  0.827\E{8}$& 5.7\\
     $\e{9}$& 293&$ 0.784\E{10}$&$  0.158\E{9}$&$  0.733\E{9}$& 6.6\\
     $\e{10}$&325&$ 0.708\E{11}$&$ 0.142\E{10}$&$ 0.882\E{10}$& 7.3\\
     $\e{11}$&358&$ 0.714\E{12}$&$ 0.149\E{11}$&$ 0.600\E{11}$& 8.2\\
     \hline
  \end{tabular}\end{center}
  \caption{Runs with the CGA on configuration A ($\kappa=10^3$).
     The last column indicates the CPU seconds taken on a \machine.}
\label{tb.cga1}\end{table}

\begin{table}[hbtp]
  \begin{center}\begin{tabular}{|c||r||c|c|c||c|}\hline
     $\ecg$ & $\ncg$ & $\ehf$ & $\egw$ &  $\ecc$ &time \\ \hline
   $\e{1}$&  23&$0.236\E{1}$&$0.394\E{1}$&$ 0.865\E{1}$&  0.5\\
   $\e{2}$& 212&$0.198\E{2}$&$0.347\E{2}$&$ 0.136\E{1}$&  4.7\\
   $\e{3}$& 335&$0.617\E{4}$&$0.114\E{3}$&$ 0.544\E{2}$&  7.4\\
   $\e{4}$& 365&$0.633\E{5}$&$0.112\E{4}$&$ 0.460\E{2}$&  8.1\\
   $\e{5}$& 397&$0.651\E{6}$&$0.124\E{5}$&$ 0.160\E{2}$&  8.8\\
   $\e{6}$& 428&$0.625\E{7}$&$0.117\E{6}$&$ 0.544\E{6}$&  9.5\\
   $\e{7}$& 459&$0.629\E{8}$&$0.116\E{7}$&$ 0.461\E{7}$& 10.2\\
   $\e{8}$& 489&$0.616\E{9}$&$0.110\E{8}$&$ 0.506\E{8}$& 10.9\\
   $\e{9}$& 517&$0.626\E{10}$&$0.109\E{9}$&$ 0.442\E{9}$& 11.6\\
   $\e{10}$&545&$0.596\E{11}$&$0.995\E{11}$&$ 0.401\E{10}$& 12.2\\
   $\e{11}$&573&$0.124\E{11}$&$0.111\E{11}$&$ 0.396\E{11}$& 12.9\\
     \hline
  \end{tabular}\end{center}
  \caption{Runs with the CGA on configuration B ($\kappa=4.4\times10^5$).
     The last column indicates the CPU seconds taken on a \machine.}
\label{tb.cga2}\end{table}

\begin{table}[hbtp]
  \begin{center}\begin{tabular}{|c||r||c|c|c||c|}\hline
     $\ecg$ & $\ncg$ & $\ehf$ & $\egw$ &  $\ecc$ &time \\ \hline
$\e1$  &  62  &$0.231\times10^{-1}$  &$0.381\times10^{-1}$  &$0.869\times10^{-1}$  & 1.4\\
$\e2$  & 642  &$0.393\times10^{-2}$  &$0.605\times10^{-2}$  &$0.153\times10^{-1}$  &14.7\\
$\e3$  & 830  &$0.310\times10^{-3}$  &$0.125\times10^{-2}$  &$0.874\times10^{-2}$  &19.2\\
$\e4$  & 863  &$0.298\times10^{-4}$  &$0.152\times10^{-5}$  &$0.691\times10^{-2}$  &20.2\\
$\e5$  & 891  &$0.287\times10^{-5}$  &$0.169\times10^{-6}$  &$0.421\times10^{-2}$  &20.7\\
$\e6$  & 922  &$0.304\times10^{-6}$  &$0.193\times10^{-7}$  &$0.325\times10^{-2}$  &21.6\\
$\e7$  & 957  &$0.318\times10^{-7}$  &$0.245\times10^{-8}$  &$0.325\times10^{-2}$  &22.4\\
$\e8$  & 987  &$0.853\times10^{-8}$  &$0.822\times10^{-8}$  &$0.231\times10^{-8}$  &23.1\\
$\e9$  &1015  &$0.766\times10^{-8}$  &$0.135\times10^{-8}$  &$0.555\times10^{-8}$  &23.8\\
     \hline
  \end{tabular}\end{center}
  \caption{Runs with the CGA on configuration C ($\kappa=3.6\times10^9$).
     The last column indicates the CPU time in seconds on a \machine.}
\label{tb.cga3}\end{table}

The first adaptive method used to compute $M^{-1/2}$ was based on
Lancz\"os algorithm \cite{borici}. In the original suggestion, the number
of Lancz\"os steps to be taken in order to reach a given precision
was investigated in terms of the variation of the eigenvalues of $M$
with the number of Lancz\"os steps. A stopping criterion {\sl \`a la}
Conjugate Gradient was proposed but its relation to the precision was
not direct. A related adaptive method based on the Conjugate Gradient
algorithm was used in \cite{ragula}. Here the stopping criterion is put
on the residual vector in the inversion of $M$. This enables a direct
control over the precision.

The CGA starts with an iteration which is almost the same as the
usual CG algorithm for the inversion of $M$---
\begin{enumerate}
\item Start from $r_1=\Phi$, $p_1=r_1$ and  $\beta_1=0$,
\item Iterate as in regular CG,
   $\alpha_i = |r_i|^2/(p_i^\dag Mp_i)$,
   $r_{i+1} = r_i-\alpha_i M p_i$,
   $\beta_{i+1} = |r_{i+1}|^2/|r_i|^2$, and
   $p_{i+1} = \beta_{i+1} p_i + r_{i+1}$.
\item Stop when $|r_{i+1}|<\ecg$.
\end{enumerate}
Note that the the only difference from the usual CG is that the vector
which is $M^{-1}\Phi$ does not need to be obtained during the
iteration.

In the  orthonormal basis of $q_i=r_i/|r_i|$, the matrix $M$ is the
composition of the matrix $Q$ whose $i$-th column is $q_i$, and a
symmetric tridiagonal matrix, $T$, 
\beq
   M=Q^\dag TQ,\quad{\rm where}\quad
      T_{ii}=\frac1{\alpha_i}+\frac{\beta_i}{\alpha_{i-1}},
	  \quad{\rm and}\quad
      T_{i,i+1}=-\frac{\sqrt{\beta_{i+1}}}{\alpha_i},
\label{diag}\eeq
where $\alpha_i$ and $\beta_i$ are defined in the iteration above.
Then compute the eigenvalues and eigenvectors of this truncated
tridiagonal matrix T,
\beq
   T=U\Lambda U^\dag
\label{trid}\eeq
where $\Lambda$ is the diagonal matrix of the eigenvalues and $U$ the
matrix of the eigenvectors in the basis $Q$.
The CGA solution is
\beq
   L[\Phi] = Q^t U \Lambda^{-1/2}U^\dag Q \Phi/|\Phi|.
\label{solve}\eeq
The adaptability of the algorithm arises from the fact that we retain
only the vectors $q_i$ which contribute significantly to the inverse of
$M$ and we stop the iterations for $i=\ncg$ when $|r_{\ncg+1}|<\ecg$.

The contribution to $M^{-1/2}$ of the smallest eigenvalue 
$1/\lambda_{\min}$ of $M$ will be only $1/\sqrt{\lambda_{\min}}$.
Since the stopping criterion $|r_{i+1}|<\ecg$ is meant
to compute $M^{-1}$ it is more stringent than required. One can
be more generous for $M^{-1/2}$, and use instead the stopping criterion
\beq
   |r_{i+1}|<\ecg/\sqrt{\lambda_0^{(i)}}
\label{stop}\eeq
where $\lambda_0^{(i)}$ is an upper bound of $\lambda_{\min}$.
Fortunately a reasonable estimate can be obtained at each iteration $i$
without large overheads. For any tridiagonal matrix $T$ of order
$\nor$, the number of eigenvalues greater than a fixed number $\mu$ is
the number of positive values of $d^{(j)}$, where this set of numbers is
defined by $d^{(1)}=T_{11}-\mu$ and
\beq
   d^{(j)}=T_{jj}-\lambda_0^{(i)}-(T_{j-1,j})^2/d^{(j-1)},
\label{next}\eeq
for $2\le j\le\nor$ \cite{book2}. An upper bound for $\lambda_0^{(i)}$
can always be fixed by searching for a number for which at least one
of $d^{(j)}$ is non-positive. This can be done by bisection, starting from
the initial estimate at the first step, $\lambda_0^{(1)}=T_{11}$. While
this procedure increases the complexity by order $\log\ncg$, the new
stopping criterion in Eq.\ (\ref{stop}) has two advantages over the
usual CG stopping criterion--- first, $\ncg$ is reduced and, second,
the method becomes better adaptable since the observed $\ehf$ for a
given $\ecg$ becomes independent of $\lambda_{min}$.

Practically, to do the computation without storing the orthonormal
basis $Q$, one makes $\ncg$ iterations to get the truncated matrix $T$,
computes the matrix $U$ and the diagonal $\Lambda$ using standard
methods \cite{golub}, and then repeats the $\ncg$ iterations to compute
the solution $L[\Phi]$.  The most stringent restriction on the algorithm
seems to be that one cannot use any pre-conditioning and must always
start the iterations from $p_1=r_1=\Phi$.  This algorithm has only one
parameter, $\ecg$. The algorithm automatically adjusts the number of
iterations to achieve the specified  precision irrespective of the
condition number.  Thus, no configuration dependent tuning of algorithmic
parameters is necessary when employing the CGA for QCD applications.

The complexity of the CGA is
\beq
   {\cal C}_{CGA} \simeq 2w\ncg V+\omega\ncg^2
      \simeq 2w'V\log\left(\frac1\ecg\right)\log\kappa
\label{compcga}\eeq
where $\omega$ is a number independent of $V$. The $\ncg^2$ term comes from
the handling of the tridiagonal matrix, and can be neglected since
$\ncg\ll V$. The space complexity is the same as that of a standard CG---
\beq
   {\cal S}_{CGA} \simeq 8N_c(N_c+3)V.
\label{spaccga}\eeq
Since the method is adaptive, no deflation is necessary. However,
deflation reduces the condition number of the matrix, and hence
could improve the complexity by reducing $\ncg$. Nevertheless, for
reasons that we have already discussed in connection with CA and ORA,
deflation is unlikely to improve the performance at fixed physics when
taking the limit of large $V$.

The results of our numerical tests for this algorithm are collected
in Tables \ref{tb.cga1}--\ref{tb.cga3}. Note that for all three test
configurations there is a threshold in $\ecg$ above which $\ecc\le10\ehf$
and below which $\ecc$ is roughly constant. The threshold value of $\ecg$
is somewhat larger than $\lambda_{\min}$ for the configuration. Similar
thresholds are also seen for the ORA and ZA. This behaviour possibly
reflects the existence of a large unconverged subspace in the CG
iterations.

\section{Comparing the algorithms}\label{sc.comp}

\begin{figure}[htbp]
   \begin{center}\scalebox{0.6}{\includegraphics{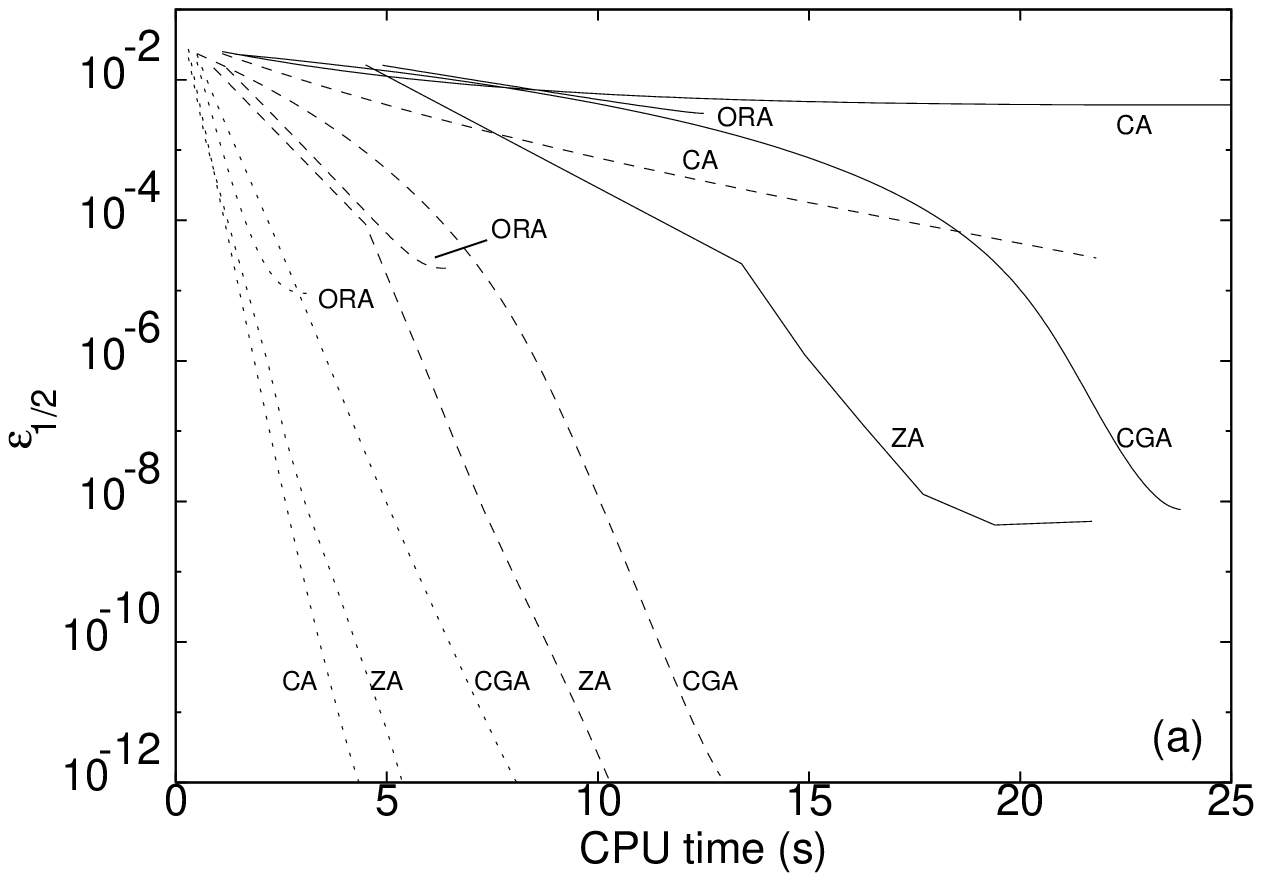}}
                 \scalebox{0.6}{\includegraphics{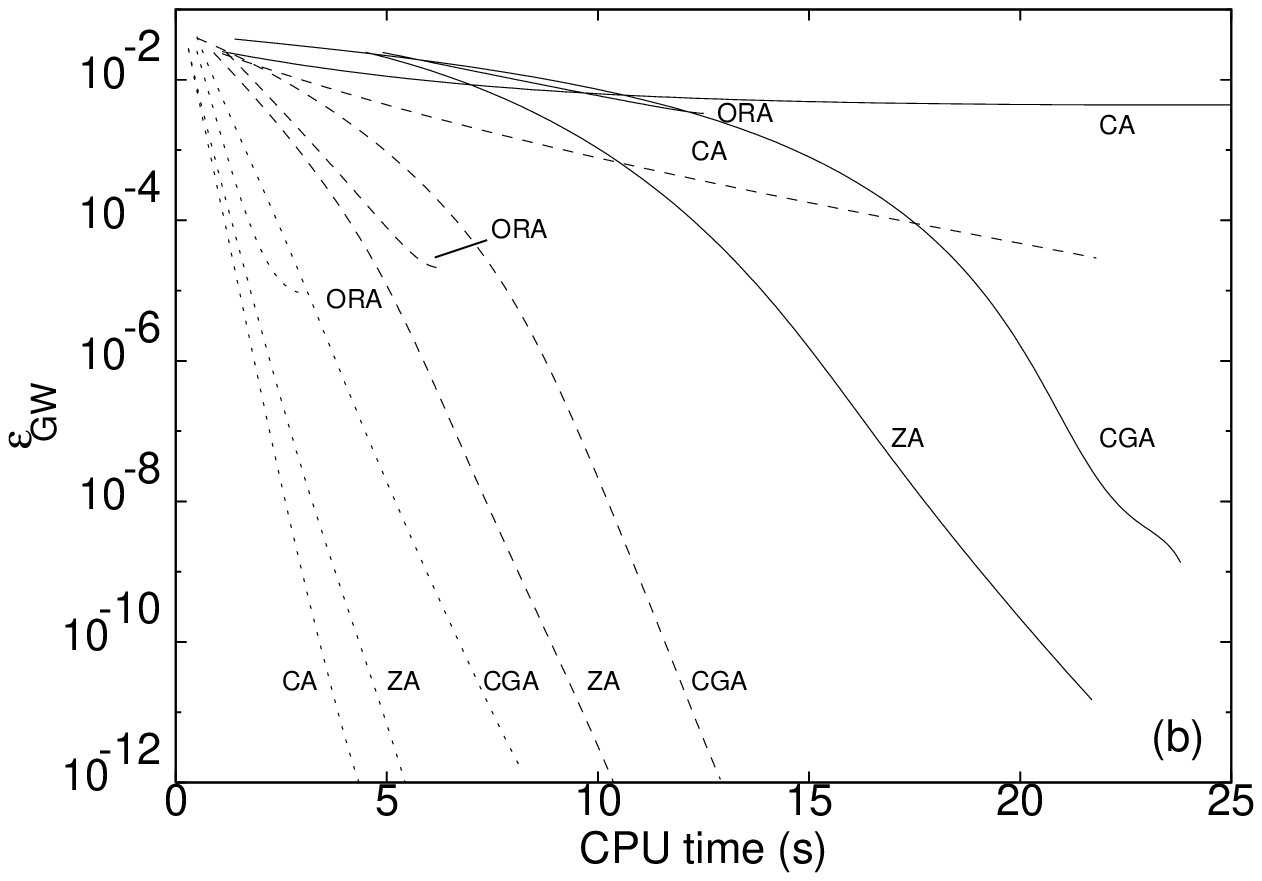}}
                 \scalebox{0.6}{\includegraphics{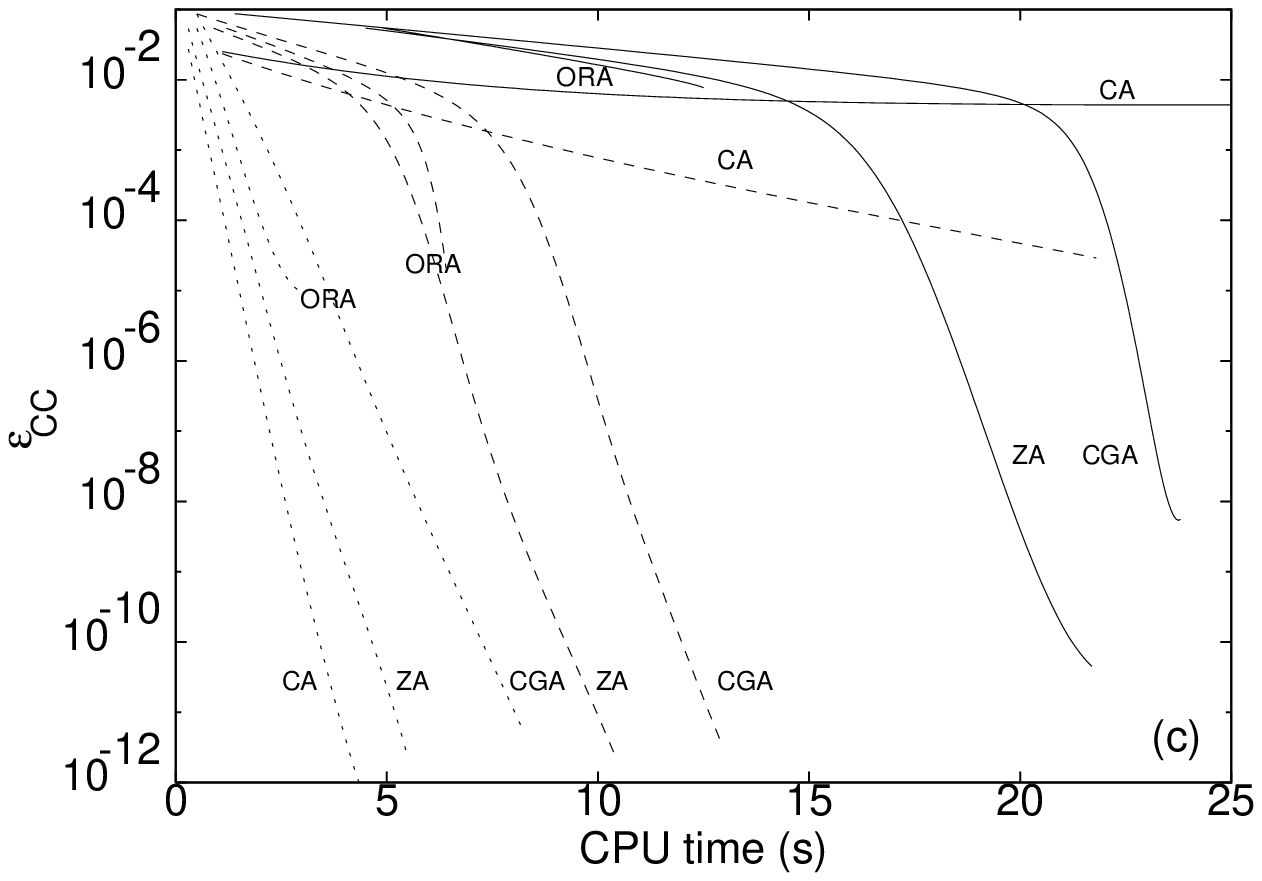}}
                 \scalebox{0.6}{\includegraphics{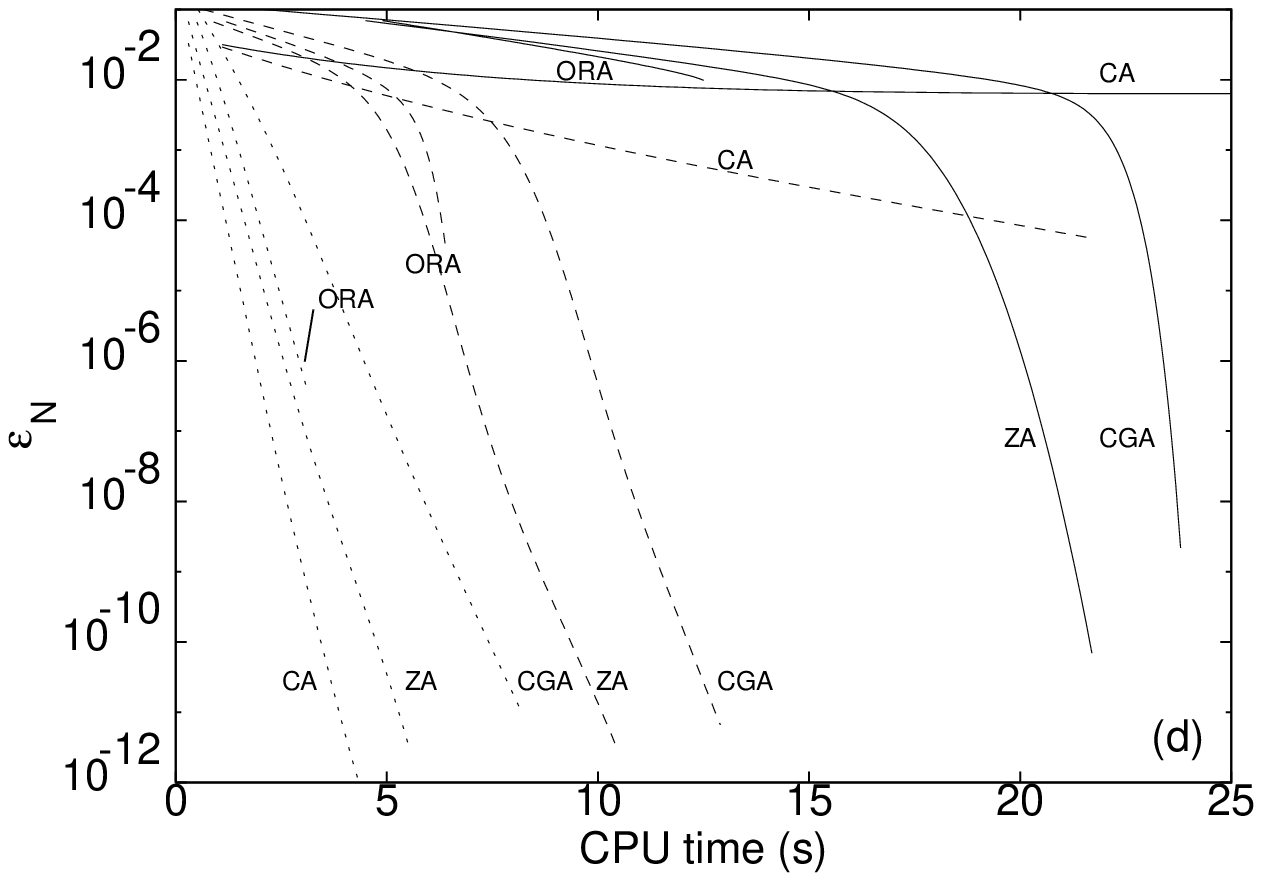}}\end{center}
   \caption{Error limits as a function of the CPU time taken on a \machine--- (a) $\ehf$, (b) $\egw$, (c) $\ecc$ and (d) $\enm$. In each
      case the dotted line is for configuration A ($\kappa=10^3$), the dashed
      line for configuration B ($\kappa=4.4\times10^5$) and the full line for
      configuration C ($\kappa=3.6\times10^9$).}
\label{fg.comp}\end{figure}

\begin{figure}[htbp]
   \begin{center}\includegraphics{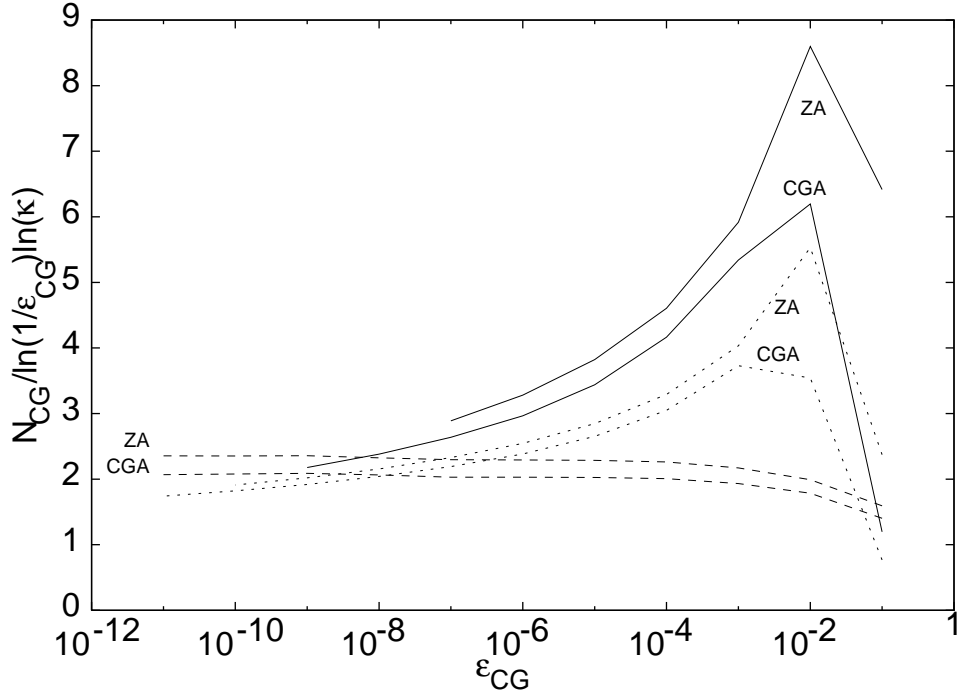}\end{center}
   \caption{The scaling of $\ncg$ with $\kappa$ and $\ecg$. These are results of
      computations with configurations A (dashed lines), B (dotted lines) and C
      (full lines).}
\label{fg.complex}\end{figure}

\begin{figure}[htbp]
   \begin{center}\includegraphics{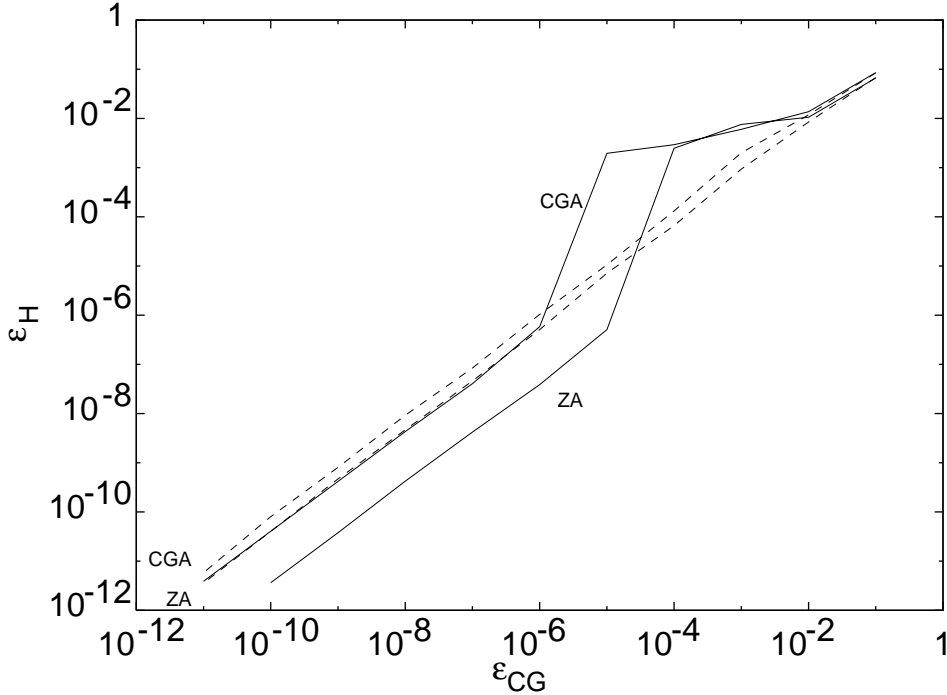}\end{center}
   \caption{The scaling of $\ehm$ with $\ecg$ in the CGA and ZA for configurations
      A (dashed lines) and B (full lines).}
\label{fg.nonlin}\end{figure}

\begin{figure}[htbp]
   \begin{center}\includegraphics[angle=270]{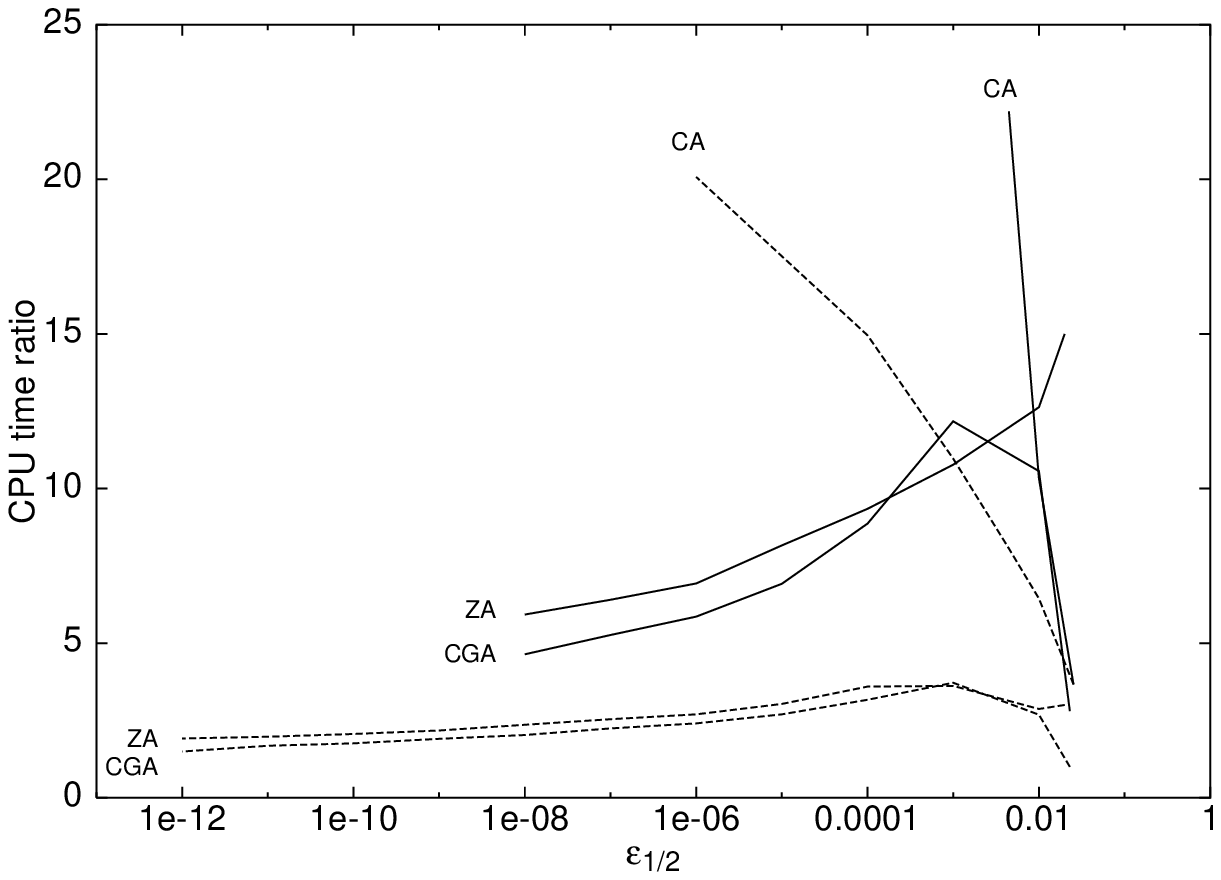}\end{center}
   \caption{The ratio of CPU time taken at fixed error, $\ehf$, for (a) configurations
      B and A (dashed lines) and (b) configurations C and A (full lines) for three
      different algorithms.}
\label{fg.adapt}\end{figure}

In Figure \ref{fg.comp} we have collected different measures of performance 
of the four algorithms we investigated in this paper, namely,
the Optimized Rational Approximation (ORA) \cite{nara}, the Zolotarev
Approximation (ZA, which is also a rational expansion) \cite{wupp,ch},
the Chebychev Approximation (CA, a polynomial expansion) \cite{hepile}
and the Conjugate Gradient Approximation (CGA, an iterative method)
\cite{ragula}.  Further details can be found in
Tables \ref{tb.cheby1}--\ref{tb.cga3}.

It is clear that for modest values of the condition number of $M$,
$\kappa\le10^3$, the CA is the preferred algorithm. This is
clear from the figure, as well as our results for the algorithmic
complexities in Eqs. (\ref{compca}), (\ref{compora}), (\ref{compza}) and
(\ref{compcga}). However, with increasing condition numbers the performance
of CA rapidly degrades. This is visible in the figure as well as in our
analysis of the adaptability in Eq.\ (\ref{adaptca}). We have argued earlier
that these drawbacks of the CA are generic to all polynomial expansions.

The ORA, in its present form with fixed $\nor$, also suffers from a lack of 
adaptability. In principle, this can be alleviated if the order of the 
approximation can be chosen adaptively. We have implemented the ZA, which is 
another rational approximation, for several different choices of order. As can
be seen from Tables \ref{tb.zolo1}- \ref{tb.zolo3}, and from Figure
\ref{fg.comp}, this improves the performance tremendously. For
comparable CPU times, corresponding to low order ZA, the performance
is at least one order better than that of ORA on all configurations.
The key to improving the performance of rational approximations 
is the automatic variation of the order $\nor$ with the condition
numbers.  In our tests we have simulated adaptability by working
with several different orders and retained the one corresponding only
to a small fraction of the inversion error.   The so-tuned order is only
slightly higher than that obtained in an automatic procedure defined
at the end of section \ref{sc.za}.

The CGA depends on only one parameter $\ecg$. For a given value of
$\ecg$, the corresponding errors $\ehf$ and $\egw$ are almost independent
of the condition number of the matrix, thanks to the relaxed stopping criterion.
The price for such a good adaptability is a computing time which is 50\%
higher than ZA for a given accuracy (70\% excess if $No$ is small, 20\% for
large $No$). The price, however, ensures that for all the configurations one
guarantees the same order of accuracy from a given value of $\ecg$ and with
a predicted value of $\egw$.

The variation in the number of conjugate gradient iterations, $\ncg$, as the
stopping criterion, $\ecg$, is changed for the three configurations is shown in
Figure \ref{fg.complex}.  The data for the ORA are not shown in the figure
because they are very similar to those of the ZA.  Note that the curves for the
CGA lie below that for the ZA (despite the shift in ZA as compared to CGA),
which is the influence of the relaxed stopping criterion discussed in Section
\ref{sc.cga}.  Ref. \cite{wupp} has devised a similar modification for ZA
which can reduce $\ncg$ in that case.
Note that our above results for ZA did not use any such modifications;
using it will further enhance the performance of ZA reported above.

We have noted in Section \ref{sc.errors} that the relations
(\ref{rel1}) between the errors are valid for those approximations to 
$M^{-1/2}$ which commute with $M$.  In particular, we noted that for the
iterative algorithms these relations become valid, provided that $\ecg$
is sufficiently small. In Figure \ref{fg.nonlin} we demonstrate this for
$\ehm$, which is expected to be zero when $\ecg$ is small enough. For the
CGA and ZA (data for the ORA are not shown because they almost coincide
with that for ZA), $\ehm$ decreases with $\ecg$. The slopes in this plot
correspond to linear decrease when $\ecg$ is sufficiently small. Clear
non-linearities are present for larger $\ecg$ when the condition number
is large. We believe that these non-linearities are due to large 
non-converged subspaces, implying a need for high accuracy. 

For fixed order algorithms the adaptability, $\cal A$, quantifies the
configuration dependence of speed. The numerical study can be used more
directly to illustrate the adaptability by studying the slowdown in going from
configuration A to B (\ie, from $\kappa=10^3$ to $4.4\times10^5$) or from A to
C ($\kappa$ changes from $10^3$ to $3.6\times10^9$). As shown in Figure
\ref{fg.adapt}, both ZA and CGA are adaptable algorithms over a wide range of
$\ehf$.  Since ZA is faster, as seen in Figure \ref{fg.comp}, it is thus
the method of choice.   Note, however, that CGA is very comparable to it,
and may be preferred for its self-tuning ability.

We emphasise that a fair test of relative performance of algorithms is
to work without deflation. First, deflation improves the performance
of each of the algorithms we have investigated. Details are given in
the sections on each algorithm. Nevertheless the algorithms based on
rational approximation seems to be less sensitive to deflation than other
ones because of the positive shifts introduced in the matrix. 
Second, since the computation of the
eigensystem of $M$, necessary to deflation, is done at finite accuracy,
it introduces extra errors. If the error in the computation of the
eigenvalue $\lambda_i$ is $\delta_i$, then the contribution to $\ehf$ is
$\delta_i/\lambda_i$. Thus, if we want to achieve a given $\ehf$, then
we must keep $|\delta_i|\le\ehf|\lambda_i|$. When the condition number
$\kappa$ increases, this criterion becomes impossible to satisfy,
leading to catastrophic loss of accuracy.

We feel it worth pointing out that deflation is only one of many possible
methods to decrease the effective condition number of the problem. Other
preconditioning methods have not been seriously explored for overlap
Fermions. The cost of accurate numerical methods seems to suggest that
numerically stable preconditioning methods will pay a big dividend in this
problem.

This work was supported by the Indo-French Centre for Promotion
of Advanced Research under project number 2104-2.

\end{document}